\documentclass[aps,pre,twocolumn]{revtex4-1}
\usepackage{graphics,color,graphicx,amsmath}

\graphicspath{{../images_finaldrafts/}}
\begin{document}

\newcommand{\mr}[1]{\mathrm{#1}}
\newcommand{\mb}[1]{\mathbf{#1}}
\newcommand{\br}[1]{\left<#1\right>}
\newcommand{\bl}[1]{\left|#1\right|}
\newcommand{\mc}[1]{\mathcal{#1}}
\newcommand{\Bi}{\ensuremath{\mc{B}_i}}
\newcommand{\Ei}{\ensuremath{\mc{E}_i}}
\newcommand{\Ai}{\ensuremath{\mc{A}_i}}
\newcommand{\Js}{\ensuremath{J_{ij}^\mr{\,sym}}}
\newcommand{\Ji}{\ensuremath{J_{ij}^\mr{\,ide}}}
\newcommand{\sups}[1]{\ensuremath{^\mr{#1}}}
\newcommand{\s}{\sigma}
\newcommand{\seti}{\{\s_\mr{i}\}}
\newcommand{\mH}{\sum_\mr{i} h_\mr{i}\s_\mr{i} +\frac{1}{2}\sum_\mr{ij} J_\mr{ij}\s_\mr{i}\s_\mr{j}}
\newcommand{\tb}[1]{\textcolor{blue}{#1}}
\newcommand{\tr}[1]{\textcolor{red}{#1}}
\newcommand{\tg}[1]{\textcolor{green}{#1}}

\title{Statistical mechanics of the US Supreme Court}
\author{Edward D. Lee,$^1$ Chase P. Broedersz,$^1$ and William Bialek$^{1,2}$}
\affiliation{$^1$Joseph Henry Laboratories of Physics,  and Lewis--Sigler Institute for Integrative Genomics, Princeton University, Princeton NJ 08544\\ $^2$Initiative for the Theoretical Sciences, The Graduate Center, City University of New York, 365 Fifth Ave., New York NY 10016}

\date{\today}

\begin{abstract}
We build simple models for the distribution of voting patterns in a group, using the Supreme Court of the United States as an example.  The least structured, or maximum entropy, model that is consistent with the observed pairwise correlations among justices' votes is equivalent to an Ising spin glass.  While all correlations (perhaps surprisingly) are positive, the effective pairwise interactions in the spin glass model have both signs, recovering some of our intuition that justices on opposite sides of the ideological spectrum should have a negative influence on one another.  Despite the competing interactions, a strong tendency toward unanimity emerges from the model, and this agrees quantitatively with the data.  The model shows that voting patterns are organized in a relatively simple  ``energy landscape,'' correctly predicts the extent to which each justice is correlated with the majority, and gives us a measure of the influence that justices exert on one another.    These results suggest that simple models, grounded in statistical physics, can capture essential features of collective decision making quantitatively, even in a complex political context.
\end{abstract}

\maketitle

 Social and political systems, almost by definition, generate collective or emergent phenomena.  It is natural to try describing these phenomena in the language of statistical mechanics \cite{castellano+al_09,SM_collect}, but it is not always clear whether this is a metaphor or a real theory within which we can make quantitative predictions.  Here we address this problem in the context of voting on the Supreme Court of the United States (SCOTUS).  While nine justices surely are not in the thermodynamic limit, we argue that there still are quantitative predictions to be made, and that models grounded in statistical physics provide the simplest account of the observed voting patterns \cite{notes}.

\begin{figure}[b!]
    \includegraphics[width=0.9\linewidth,clip=true,trim=0 5 0 5]{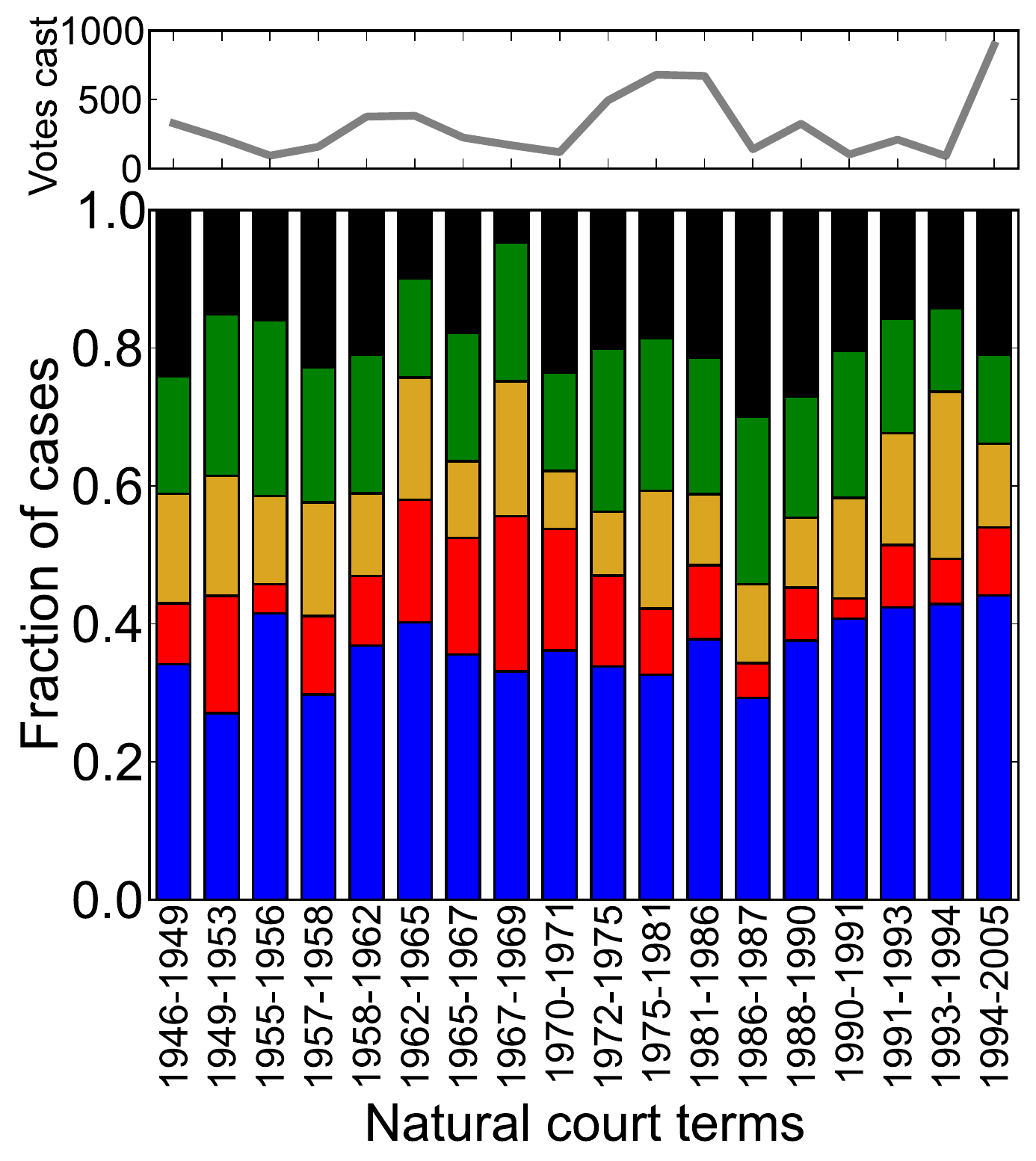}
    \caption{Distribution of dissenting votes for  natural courts that decided more than 100 cases.  The colored portions represent the number of dissenting votes: blue (0), red (1), yellow (2), green (3), black (4). On average, 36\% of votes are unanimous over the 18 natural courts shown. Above, the number of  votes cast by each natural Court.  Data from Ref \cite{data}.  \label{consensus}}
\end{figure}

SCOTUS is the highest court in the US government, consisting of nine justices  who vote on the constitutionality of legislative and executive actions. We consider natural courts, periods of time during which the membership stays constant, and focus on the second Rehnquist Court (1994--2004, $N = 895$ votes), which provides the largest data set \cite{data, initials}. The Court issues majority and minority opinions,  and these can be supplemented with other opinions; although opinions can be nuanced, each justice casts a yes ($\sigma_{\rm i} = +1$) or no ($\sigma_{\rm i} = -1$) vote, and the majority of votes decides the fate of the case.

A widely discussed fact of current political life in the United States is the strong polarization along party lines, so that consensus and unanimity have become rare.  Comments on the nature of decision making on the Supreme Court also point to strong ideological divisions between right and left, with one or two justices providing ``swing votes'' \cite{Martin2004}.  In reality, unanimous decisions are much more likely than 5-4 splits  \cite{sirovich2003}, as shown in Fig \ref{consensus}. This pattern is consistent across more than fifty years, and there is little indication that the unanimous cases are in a special class of ``easy'' decisions \cite{easy}.

\begin{figure}[tb]
    \includegraphics[width=.8\linewidth,clip=true,trim=0 32 0 55]{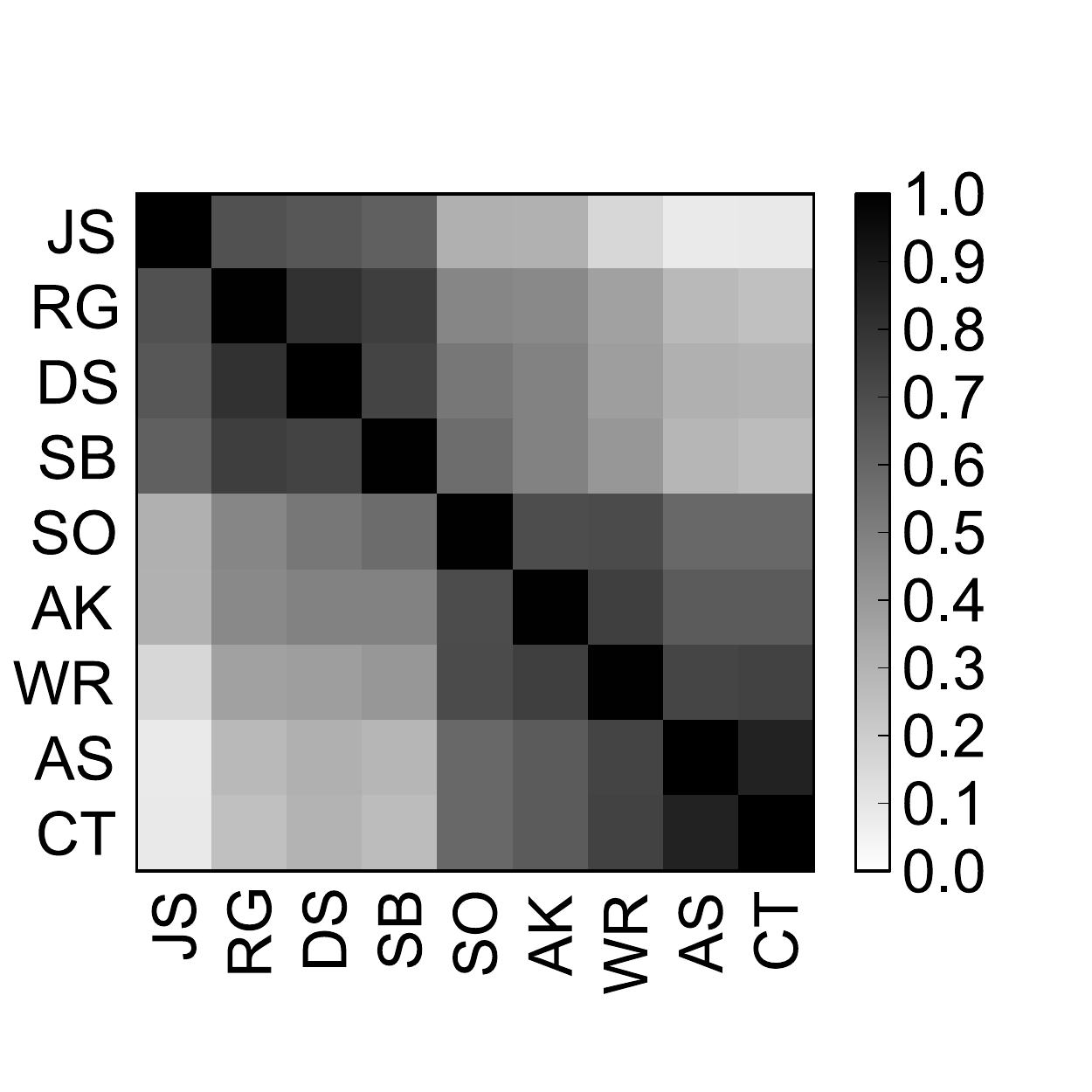}
    \caption{Correlation matrix of votes in the Rehnquist court  (1994--2005, $N = 895$ votes).  As explained in the text, yes/no votes are represented as binary variables $\sigma_{\rm i} = \pm1$ for each justice $\rm i$, and we plot $C_{\rm ij} = \langle \sigma_{\rm i}\sigma_{\rm j}\rangle$; by convention,  $\langle\sigma_{\rm i}\rangle =0$ for all $\rm i$, and the diagonal elements $C_{\rm ii} =1$.  Individual justices are identified by their initials \cite{initials}, ordered roughly from ideological left (JS) to right (CT) \cite{data}. Note that all correlations are positive, despite ideological differences. The standard error in estimating $C_{\rm ij}$ is given by $\delta C_{\rm ij} = [(1-C_{\rm ij}^2)/N]^{1/2}$; with $N = 895$ we have $\delta C_{\rm ij} < 0.034$ for all $\rm ij$.\label{Cij}}
\end{figure}

The definition of yes and no in each case is determined by decisions in lower courts, and thus is somewhat arbitrary.  As a start, we imagine that the opposite definition was also possible, so that the voting patterns $\{\sigma_{\rm i}\}$ and $\{-\sigma_{\rm i}\}$ are equally likely, and we return to this problem below.  With this symmetry, we are guaranteed that the average vote is neutral,  $\langle \sigma_{\rm i}\rangle = 0$ for all justices.  Then the first nontrivial voting statistic is the matrix of correlations, $C_{\rm ij} = \langle \sigma_{\rm i}\sigma_{\rm j}\rangle$, shown in Fig \ref{Cij}.  While one might have expected that justices known to have opposite ideological positions would tend to cast opposing votes, we see that all correlations  are positive.  We would like to understand not just these pairwise correlations, but the entire distribution of voting patterns $P(\{\sigma_{\rm i}\})$.

There are an infinite number of distributions $P(\{\sigma_{\rm i}\})$ that are consistent with the observed correlations $C_{\rm ij}$.  Out of all these models, we can ask for the one which has the least structure, or equivalently the one that generates the most random patterns of votes.  This distribution embodies the {\em minimal} implications of the pairwise correlations, and involves no further assumptions.  We know from Shannon \cite{shannon_48,cover+thomas_91} that the qualitative concepts of ``most random'' and ``least structured'' have a unique formalization, namely that we should search for the distribution that has the maximal entropy consistent with $C_{\rm ij}$ \cite{jaynes_57}.
Thus, we wish to construct $P(\{\sigma_{\rm i}\})$ that has the largest value of the entropy
\begin{equation}
S[P(\{\sigma_{\rm i}\})] \equiv -\sum_{\{\sigma_{\rm i}\}} P(\{\sigma_{\rm i}\}) \ln P(\{\sigma_{\rm i}\}) ,\label{ent1}
\end{equation}
while insisting that the correlations match  their experimental values,
\begin{equation}
\sum_{\{\sigma_{\rm i}\}} P(\{\sigma_{\rm i}\}) \sigma_{\rm k}\sigma_{\rm j} = C_{\rm kj}.
\label{eq:Cij}
\end{equation}
The solution to this constrained optimization problem is a Boltzmann--like distribution,
\begin{equation}
P(\{\sigma_{\rm i}\}) = {1\over Z} e^{-E(\{\sigma_{\rm i}\})} , \label{ising1}
\end{equation}
where the effective energy of each state is given by
\begin{equation}
E(\{\sigma_{\rm i}\}) = -{1\over 2} \sum_{{\rm i}\neq{\rm j}} J_{\rm ij} \sigma_{\rm i} \sigma_{\rm j},\label{ising2}
\end{equation}
and the parameters $J_{\rm ij}$ must be adjusted to satisfy the constraints in Eq (\ref{eq:Cij}); as usual the partition function serves to normalize the distribution, so that
\begin{equation}
Z = \sum_{\{\sigma_{\rm i}\}}e^{-E(\{\sigma_{\rm i}\})}.
\label{ising3}
\end{equation}
We recognize Eqs (\ref{ising1}--\ref{ising3}) as mathematically equivalent to the Ising model of a magnet, with ``spins'' $\sigma_{\rm i}$ interacting through  ``couplings''  $J_{\rm ij}$.
 
\begin{figure}[t]
 \includegraphics[width=.8\linewidth,clip=true,trim=0 32 0 55]{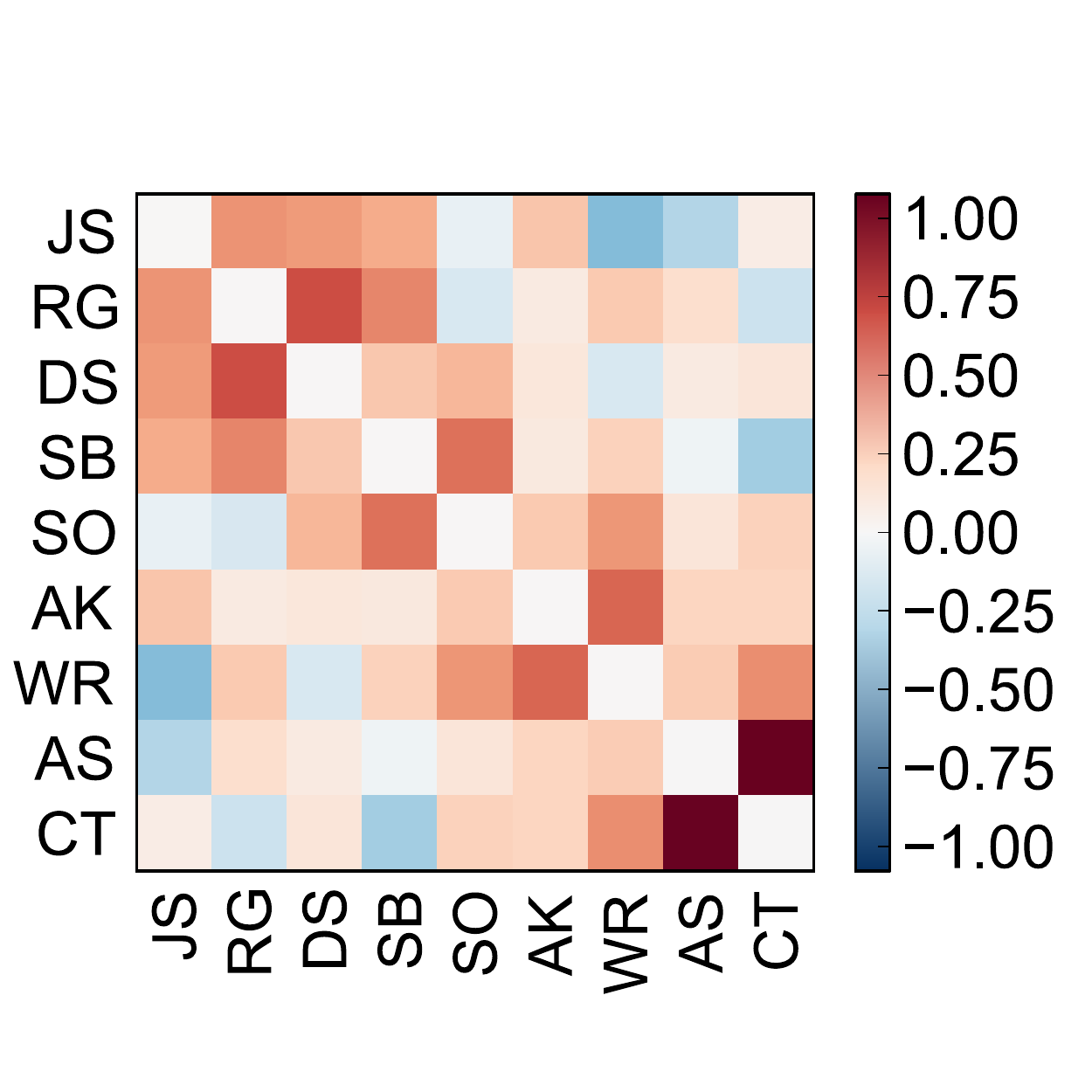}
\caption{Effective interactions in the Rehnquist court. We show the couplings $J_\mr{ij}$, as in Eq (\ref{ising2}).  Some $J_\mr{ij}$ are negative despite all positive $C_\mr{ij}$ in Fig \ref{Cij}. For a discussion of errors in the estimates of $J_{\rm ij}$, see Appendix \ref{sec: solving inverse}.\label{Jij}}
\end{figure}

With $N=9$ justices, Eq (\ref{eq:Cij}) provides 36 simultaneous nonlinear equations for the $J_{\rm ij}$, and it is straightforward to solve these numerically. The result for the Rehnquist court (Fig \ref{Cij}) is shown in Fig \ref{Jij}.  The first thing we notice is that the interactions $J_{\rm ij}$, in contrast to the correlations $C_{\rm ij}$, are both positive and negative.

\begin{figure*}[tb]
  \includegraphics[width=\linewidth,clip=true,trim=0 5 0 0]{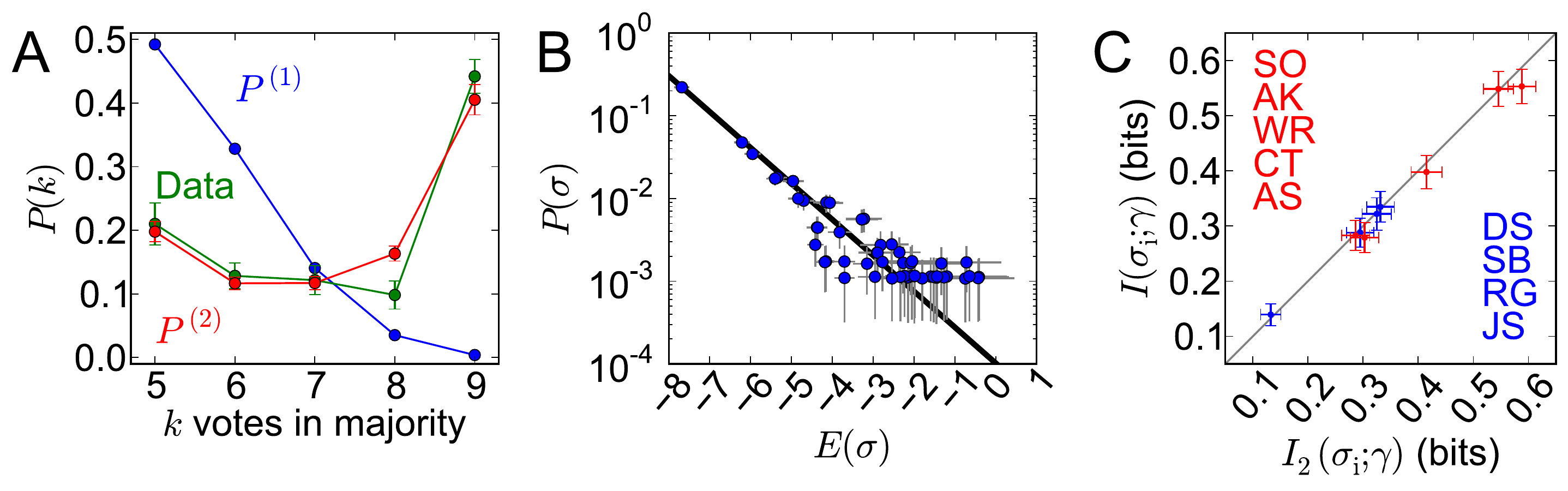}
  \caption{Testing the maximum entropy model for the Rehnquist court. (A) Probability of $k$ votes in the majority.  We compare the data (green) with the predictions of the pairwise maximum entropy model $P^{\,(2)}$ (red), and with a model of independent votes $P^{\,(1)}$ (blue). (B) Probability of each of the 210 observed voting patterns ${\mathbf\sigma}\equiv \{\sigma_{\rm i}\}$ vs the ``energy'' in Eq (\ref{ising2}); line is Eq (\ref{ising1}). Errors in probability arise, as usual, from counting; errors in the energy are propagated from errors in estimating the parameters $J_{\rm ij}$. Only states that appear more than once are shown, setting a floor for $P(\sigma )$. (C) Mutual information $I(\sigma_{\rm i};\gamma)$ between individual votes $\sigma_{\rm i}$ and the decision $\gamma$ of the majority, compared with $I_2(\sigma_{\rm i}; \gamma)$ from the model.  Conservatives are red and liberals blue, from highest $I(\sigma_{\rm i};\gamma)$ to lowest according to data (Table \ref{influence}). \label{fit}}
\end{figure*}

The correlation matrix tells us that a positive vote by the most conservative justice (CT) raises the probability of a positive vote by the most liberal justice (JS) by $C_{\rm ij}/2 \sim 4\%$, but this includes all the indirect paths through other members of the court for these justices to influence one another. In the context of the joint distribution for all votes, a positive vote by CT, with all other votes held fixed, contributes a factor $e^{J_{\rm ij}} \sim 1.1$  to the probability of a positive vote by JS, surprisingly pulling JS further in the same direction, at odds with the ideological intuition. But another ideological opposite like AS, with a very similar voting record to CT, contributes a factor of $e^{J_\mr{ij}} \sim 0.7$, \emph{decreasing} the probability by 30\%. Thus, this model constructed only from  (measured) positive correlations unmasks the hidden negative interactions, but shows that these do not conform fully to the binary intuition of negative influence spanning the ideological spectrum and positive influence within blocs.

Before proceeding, we address the errors in our estimates of the $J_{\rm ij}$.    Individual $J_{\rm ij}$ are determined with standard deviations $\Sigma_J = 0.07 - 0.2$, but these errors are correlated.  What really matters is our ability to predict the ``energy'' of each state through Eq (\ref{ising2}), and we find that for low energy states the errors in the energy are $\Sigma_E = 0.2 - 0.3$.  Thus, we can determine the parameters of the model well enough  to predict probabilities of common states (those which occur several times in the data) with an accuracy of $\sim 30\%$.  For details see Appendix \ref{sec: solving inverse}.

Just because the maximum entropy model is the least structured model consistent with the observed pairwise correlations does not mean that it is correct, since we can imagine interactions among groups of justices that would not be captured by measurements on pairs \cite{Epstein}.  The model, however, predicts the full distribution over voting patterns, and thus can be tested in various ways.  First, we can calculate the probability that the vote is split $(k, 9-k)$, with $k = 5, 6, \cdots , 9$ votes in the majority, shown in Fig \ref{fit}A.  While there are small quantitative discrepancies, the model correctly predicts that unanimous votes are twice as likely as 5--4 splits, reproducing all the probabilities with $\sim 10\%$ accuracy.   Note that if the votes of the individual justices were independent, then unanimous votes would occur only $\sim 1\%$ of the time, while 5--4 splits would be the most common outcome.  The observed tendency toward unanimity  is described by the model as a truly emergent phenomenon, the minimal consequence of the observed correlations among pairs of justices.

Although error bars are larger, a second test is to estimate the probability of every voting pattern in the data and compare these estimates with the predictions of the model.  This is shown in Fig \ref{fit}B, and we see that theory and experiment agree within error bars for almost all patterns that occur more than once in the data.

Deviations between the model and the data are small, but could add up to significant effects.  A third test, then, is to compare mutual information between the votes of individual justices and the majority vote $\gamma$, as shown in Fig \ref{fit}C. The values of the mutual information $I(\sigma_{\rm i}; \gamma)$ range over a factor of four, so that SO's vote provides $0.55\pm 0.06$ bits of information about the decision of the court as a whole, while JS's vote provides only $0.14\pm 0.04$ bits.  This pattern, related to previous observations   \cite{Black1948, Martin2004}, is reproduced very accurately by the model.

Taken together, the three results in Fig \ref{fit} provide strong evidence that our model for the distribution of voting patterns captures the interesting structure in these data.  We emphasize that this model is built only from measured pairwise correlations, and that once we have found the maximum entropy model there is no fitting of the data in Fig \ref{fit}; instead we have unambiguous, quantitative predictions, with no adjustable parameters.

The most direct test of maximum entropy models is to measure the entropy itself.  If we build maximum entropy models that capture correlations of order $n$, then we generate a sequence of models with strictly decreasing entropy, $S_1 > S_2 > S_3 > \cdots > S_K$ \cite{schneidman+al_03}. In this sequence, $n=1$ corresponds to a model of independent voting by each justice, while $n=K =9$ corresponds to the exact model which reproduces correlations of all orders.  The total amount of correlation in the system can be measured by the multi--information, $I_K = S_1 - S_K$ \cite{ent_est}.  The pairwise maximum entropy models capture a fraction $F = (S_1 - S_2) /I_N$ of this structure.  Over all the natural courts shown in Fig \ref{consensus}, we find that $F = 0.95\pm 0.03$.

The energy function $E(\{\sigma_{\rm i}\})$ in Eq (\ref{ising2}) includes competing interactions, since the $J_{\rm ij}$ have both positive and negative signs.  We expect  that this competition will generate multiple local minima in the energy landscape \cite{mezard+al_87}, or local maxima in the probability distribution,  where by a local maximum we mean that flipping the vote of any single justice lowers the probability of the voting pattern.  For the Rehnquist court, with $J_{\rm ij}$ in Fig \ref{Jij}, we find that more than 99\% (508/512) of the patterns fall into just $2\times 3$ ``valleys'' in the energy landscape.  The most populated pair of valleys are built around the two possible unanimous votes.  A second pair of valleys are built around 5-4 splits that occur precisely along ideological lines (WR, SO, AS, AK, and CT vs. JS, DS, RG, and SB).  The third pair of valleys have at their base 7-2 splits, in which the most conservative and tightly correlated justices (AS and CT) dissent from the majority.   Essentially all possible voting patterns thus are organized around intuitively understandable, prototypical patterns.  Importantly, this structure of coalitions among multiple justices emerges from the pairwise maximum entropy model with no additional assumptions; for more details see Appendix \ref{app:landscape}.  The organization of the energy landscape is truly emergent, since the pairwise interactions among the justices (Fig \ref{Jij}) do not have a rigid block structure that would support the 5--4 ideological splits.

A basic question about the dynamics of a court concerns the influence that individual justices have on the majority decision.  One way to measure this is by the mutual information $I(\sigma_{\rm i}; \gamma)$ (Fig \ref{fit}C); because the votes are symmetric binary variables, this is equivalent  to measuring the correlation $c_{\rm i} = \langle \gamma\sigma_{\rm i}\rangle - \br{\gamma}\br{\sigma_{\rm i}}$ \cite{mutual info}. We can exploit the mapping of our model onto a system of spins and imagine what happens if we add to the energy function a term $\sum_{\rm i}h_{\rm i}\sigma_{\rm i}$, such that each justice's vote is biased by a small ``magnetic field'' $h_{\rm i}$.  Then it is natural to ask how the bias of one justice propagates to the majority, that is $\chi_{\gamma{\rm i}} = \partial\langle\gamma\rangle /\partial h_{\rm i}$.  But because our model is equivalent to an equilibrium statistical mechanics problem, we have  $\chi_{\gamma{\rm i}} = c_{\rm i}$.  Thus,  seemingly different ways of measuring influence are in fact the same, and none succeeds in isolating the direct effect of one justice on the majority.

In the analogy to a magnet, each justice experiences an effective field $h_{\rm i}^{\rm eff} = \sum_{{\rm j}\neq{\rm i}} J_{\rm ij}\sigma_{\rm j}$ from the other justices. If we imagine that  justice $\rm i$ can ``lean'' in a positive direction by an amount $\epsilon$, this creates fields $\Delta h_{{\rm j}\neq{\rm i} }^{\rm eff} ({\rm i})= J_{\rm ji}\epsilon$.  But through feedback, these fields will also bias the vote of justice $\rm i$.  To isolate the influence of this one justice, we add an additional field, $\Delta h_{\rm i}^{\rm eff}({\rm i}) =   -(\epsilon/\chi_{\rm ii}) \sum_{\rm j} \chi_{\rm ji} J_{\rm ij}$, that serves to hold fixed the average vote of justice $\rm i$, where $\chi_{\rm ji} = \partial\br{\s_\mr{j}}/\partial h_\mr{i}$.  Now we can ask how the average majority vote would change if justice $\rm i$ sends a signal indicating an $\epsilon$ tendency toward  a positive vote, but does not actually cast this vote.  The resulting susceptibilities, $\Gamma_{\rm i} = (1/\epsilon)\sum_{\rm j} \chi_{\gamma{\rm j}}\Delta h_{\rm j}^{\rm eff}({\rm i})$,  are summarized in Table \ref{influence}.

\begin{table}[tb]
      \begin{tabular}{|c|c|c|}
      \hline
          & $I(\sigma_{\rm i}, \gamma)$ & $\Gamma_{\rm i}$ \\
    \hline
    JS    & $0.14 \pm 0.02$   & $0.13 \pm 0.02$ \\
    RG    & $0.29 \pm 0.03$   & $0.25 \pm 0.02$ \\
    DS    & $0.33 \pm 0.03$   & $0.30 \pm 0.02$ \\
    SB    & $0.32 \pm 0.03$   & $0.28 \pm 0.02$ \\
    SO    & $0.55 \pm 0.03$   & $0.34 \pm 0.01$ \\
    AK    & $0.55 \pm 0.03$   & $0.35 \pm 0.01$ \\
    WR    & $0.40 \pm 0.03$   & $0.23 \pm 0.02$ \\
    AS    & $0.28 \pm 0.03$   & $0.20 \pm 0.03$ \\
    CT    & $0.28 \pm 0.03$   & $0.14 \pm 0.02$ \\
    \hline
    \end{tabular}
\caption{Measures of influence. Mutual information $I(\sigma_{\rm i}, \gamma)$ between the vote of Justice $\sigma_{\rm i}$ and of the Court $\gamma$,  as in Fig \ref{fit}C. Influence increases as we move from ideological extremes into the medians. Susceptibility $\Gamma_{\rm i}$ of the majority to a signal from justice $\rm i$, as defined in the text.  }\label{influence}
\end{table}%

Using the susceptibility $\Gamma_{\rm i}$ as a measure of influence, we see that influence tends to increase as we move from the ideological extremes into the medians. The ``median Justices'' SO and AK are traditionally viewed as swing voters and indeed have maximal influence \cite{Martin2004}. At both ideological extremes, we see that Justices JS and CT have minimal---and nearly identical---influence. Interestingly, the Chief Justice WR has a nearly median ranking according to this measure of individual influence, although his votes are more strongly predictive of the majority.
 
It is perhaps surprising that we have been able to build accurate models without accounting for justices' ideological biases.  On the contrary, structures that embody these biases emerge from the model.  Suppose, however, that we go back to the beginning and identify yes/no votes  with right/left political positions \cite{ideology}.   Because we have lost the symmetry between yes and no, we want to build a maximum entropy model that matches both the pairwise correlations as before and the expectation values of the votes from individual justices, $\langle\sigma_{\rm i}\rangle$.  This model still has the Boltzmann form of Eq (\ref{ising1}), but now the energy function has explicit fields $h_{\rm i}$ acting on each spin. One might expect that this is a very different model.  In particular, it could be that the correlations between the votes of different justices are merely the reflection of their ideological biases, so that if we keep track of these, all the interactions $J_{\rm ij}$ will vanish. In fact, the couplings $J_{\rm ij}$ in the ideological model  are almost the same as in the symmetrized model, with a correlation coefficient of 0.98. This means, for example, that the correlations between votes by AS and JS (discussed above) arise not merely because they adhere to opposite biases, but because the genuinely tend to vote against one another. For a more detailed analysis, see Appendix \ref {sec:ideological_votes} and Ref \cite{thesis}.

To summarize, a pairwise maximum entropy model is sufficient to capture the voting distribution of SCOTUS.   Although there are small deviations from the predictions of the model, its success suggests that simple models, grounded in statistical physics, can provide surprisingly accurate descriptions of collective behavior even in a complex, political context.    Importantly, the notion that voting patterns---here, the competition between unanimity and ideological division---emerge from interactions among the justices in the same way that, for example, magnetism emerges from interaction among spins is {\em not} a metaphor.  Rather than being ``like a magnet,'' the voting patterns are mathematically equivalent to the spin configuration of a very particular Ising magnet, whose interactions are determined by the available data \cite{connect,bialek+al_12,tkacik+al_13}.

\begin{acknowledgments}
We thank G Berman, M Castellana, B Daniels,  J Flack, D Krakauer, and M Tikhonov  for many helpful discussions. Work in Princeton was supported in part by the National Science Foundation through Grants PHY--0957573 and CCF--0939370, by the WM Keck Foundation, and by the Lewis--Sigler Fellowship.  Work at CUNY was supported in part by the Burroughs Wellcome Fund and by the Winston Foundation.
\end{acknowledgments}

\section*{Appendix}

\appendix

The goal of this Appendix is two fold.  First, there are a variety of technical issues which should be clarified.  Second,  we would like to make the relevant ideas as accessible as possible, beyond a physics audience.  Thus, we give more than the usual background.

\medskip

\section{The data}

\begin{figure}[tb]
 \includegraphics[width=0.8\linewidth]{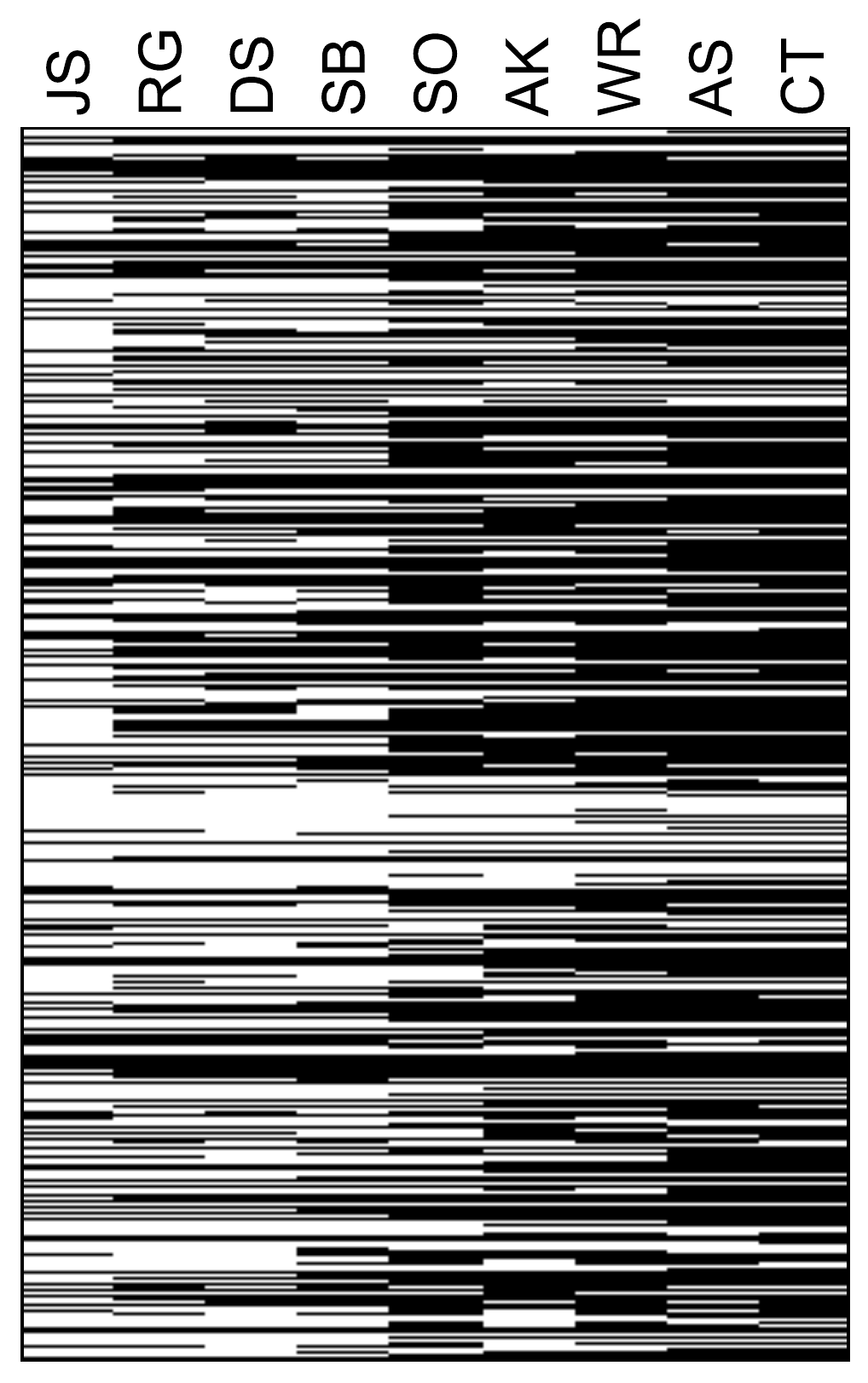}
\caption{Raw data on the Rehnquist court \cite{data}. Black denotes a positive ($\sigma_{\rm i} = +1$) vote, and white a negative ($\sigma_{\rm i} = -1$) vote.  As explained in the text, the sign is set along an ideological scale so that positive (negative) corresponds to a conservative (liberal) decision as defined by \cite{data}.} \label{rawdata}
\end{figure}

Just ten years ago, Sirovich's pioneering analysis of voting patterns on the Supreme Court required the manual entry and coding of data from individual cases \cite{sirovich2003}.  Our task has been made much easier by the efforts of Spaeth and coworkers, who have compiled a large body of data and made it accessible through a web site at Washington University in St Louis \cite{data}.  To illustrate, we show in Fig \ref{rawdata} the raw data on the Rehnquist court that forms the basis for most of our analyses. This example also shows the ideologically labeled liberal vs. conservative votes which Spaeth et al. assigned using the criteria they detail on their webpage (see also Appendix F).

\begin{figure}[tb]\centering
  \includegraphics[width=.8\linewidth]{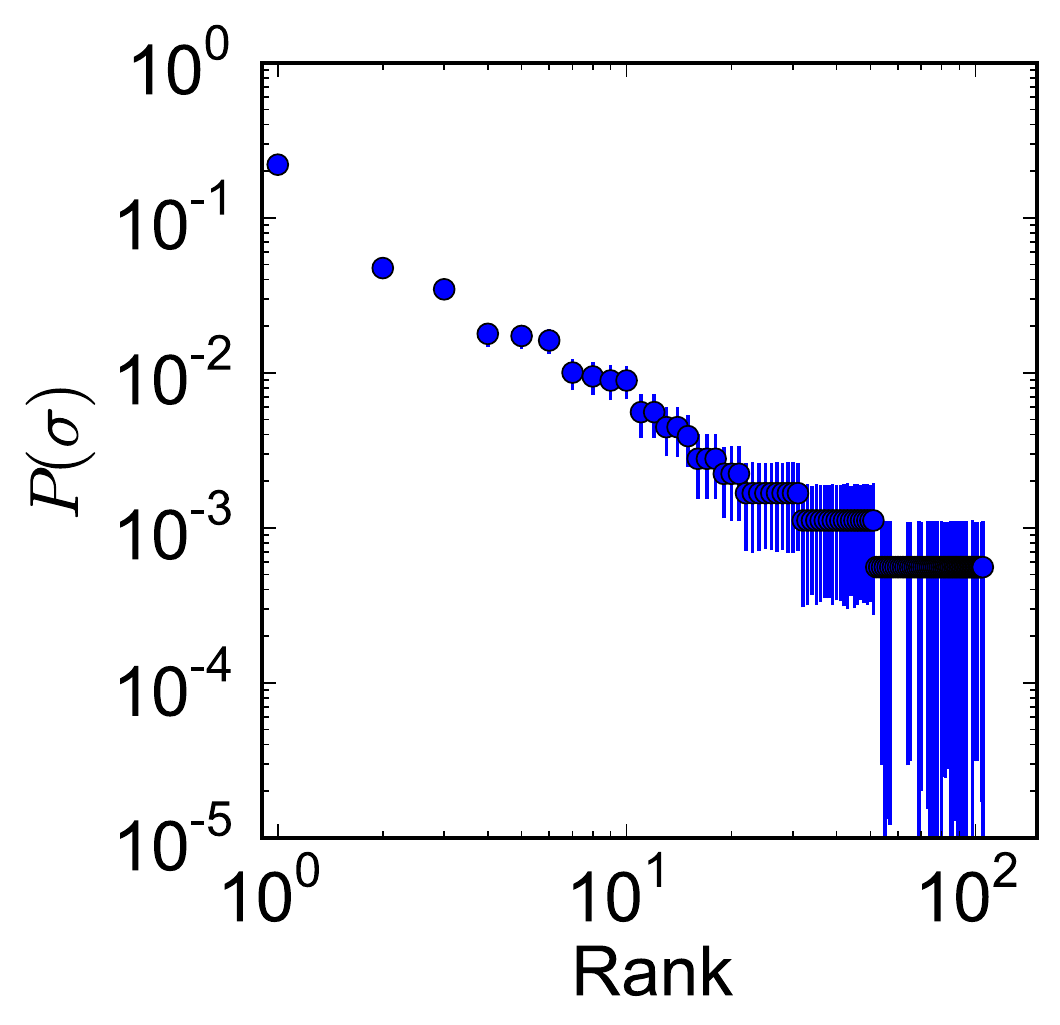}
  \caption{Zipf plot of symmetrized voting patterns in second Rehnquist Court.   Error bars are standard deviations over bootstrap samples.\label{graph:zipf}}
\end{figure}

As discussed in the main text,   the definitions of yes and no are, to some extent, arbitrary.  Much of our discussion, then, is for a symmetrized data set in which we take our samples to consist both of the observed voting patterns $\{\sigma_{\rm i}\}$ and the inverted patterns $\{-\sigma_{\rm i}\}$.  In Fig \ref{graph:zipf} we show the probability of these patterns, sorted by their rank.  This representation of the data often is called a ``Zipf plot,'' after Zipf's discussion of the distribution of words in English  \cite{zipf}.   As with words, we see that the probability has an approximately power--law dependence on rank (a straight line on the log--log plot of Fig \ref{graph:zipf}), although in the present case this is true only over a very limited dynamic range.  Although we have shown results primarily from the second Rehnquist Court, we have considered a total of 18 natural courts  available in the data set;  aspects of these different courts are discussed below.
 
\section{Maximum entropy models}

The concept of entropy has its roots in thermodynamics, roughly 150 years ago.  The idea that a maximum entropy principle could be used to build models of systems well outside the domain of thermodynamics is a more recent development, but still is more than 50 years old \cite{jaynes_57}.  The intuition, which has surprising consequences, is that we would like to make models that match certain experimental observations, but we would like to do this in a way that does not introduce any structure beyond that which is necessary to match the data.  Here we follow the  classical development of this idea, leading to Eqs (\ref{ising1}, \ref{ising2}).  For a textbook account see Appendix A.7 of Ref \cite{bialek_12}.

To be more concrete, we imagine that the system we are studying has several degrees of freedom, each described by a variable $\sigma_{\rm i}$; here these variables are the votes cast by the ${\rm i} = 1,\, 2,\, \cdots ,\, N =9$ justices.  The state of the entire system then is defined by the set of variables $\{\sigma_{\rm i}\}$, and in the simplest case what we mean by ``making a model'' is writing down the probability distribution out of which these states are being drawn, $P(\{\sigma_{\rm i}\})$.

Once we adopt a probabilistic description, experimental observations are averages, or expectation values in this distribution.  Thus, we might want to know the average vote of each justice, and this is given by
\begin{equation}\label{eq: match corr}
\langle \sigma_{\rm i}\rangle_P \equiv \sum_{\{\sigma_{\rm j}\}}\sigma_{\rm i} P(\{\sigma_{\rm j}\}) ,
\end{equation}
where the sum is over all possible patterns of votes by all the justices, and the subscript reminds us that this average is being computed in the distribution $P$.  Similarly, we might want to know the correlations between votes cast by different justices, and the pairwise correlations are given by
\begin{equation}
 \langle \sigma_{\rm i} \sigma_{\rm j}\rangle_P \equiv \sum_{\{\sigma_{\rm k}\}}\sigma_{\rm i}\sigma_{\rm j} P(\{\sigma_{\rm k}\}).
\end{equation}
These restrictions are equivalent to fixing the covariances, which be get by subtracting off the means: $C_{\rm ij} = \br{\s_\mr{i}\s_\mr{j}} -\br{\s_\mr{i}}\br{\s_\mr{j}}$. For the symmetrized data, we have that $\br{\s_\mr{i}} = 0$ so matching the pairwise correlations is the same as matching the covariances.

If we want our model to match experimental observations on these expectation values, we insist that $\langle \sigma_{\rm i}\rangle$ computed from the distribution $P(\{\sigma_{\rm i}\})$  be the same as the average computed from the experimental data,
\begin{equation}
\langle \sigma_{\rm i}\rangle_P = \langle \sigma_{\rm i}\rangle_{\rm expt} ,
\label{match1}
\end{equation}
and similarly for  the correlations,
\begin{equation}
 \langle \sigma_{\rm i} \sigma_{\rm j}\rangle_P \equiv  \langle \sigma_{\rm i} \sigma_{\rm j}\rangle_{\rm expt}.
 \label{match2}
\end{equation}
Notice that we could also make other choices, matching different features of the data.  The average votes and their pairwise correlations, however, seem like natural choices.

As emphasized in the text, Eqs (\ref{match1}, \ref{match2}) do not specify the distribution $P(\{\sigma_{\rm i}\})$ uniquely.  Indeed, there are infinitely many distributions that are consistent with these experimental observations.  Out of these infinitely many possibilities, we would like to discipline ourselves, and not introduce any structure that is not actually needed in order to match the data, where ``match the data'' now has the concrete meaning of satisfying Eqs (\ref{match1}, \ref{match2}).  Another way of saying this is that we would like a probability distribution such that, when we choose states out of this distribution, these states look as random as possible while still matching the data.

The idea that a distribution has minimal structure, or generates states with maximum randomness, might seem hopelessly qualitative.  But in 1948 Shannon proved that there is a unique way to translate this intuition into mathematical terms if we adopt some simple requirements \cite{shannon_48,cover+thomas_91}.  The result is that the only consistent measure of the randomness of states, or the lack of structure, is given by the entropy of the probability distribution,
\begin{equation}\label{def: entropy}
    S[P(\seti)] \equiv -\sum_{\seti} P(\seti) \ln P(\seti) .
\end{equation}
There is an ambiguity of units; indeed, chemists and physicists typically choose different units for the entropy even in the thermodynamic context.  The ambiguity of units is equivalent to an arbitrariness in choosing the base of the logarithm.  Here we choose the natural log, but in other settings it is conventional to choose the logarithm base 2, in which case the units of entropy are bits.  Importantly, the entropy which Shannon found as a measure of randomness or disorder in probability distributions is {\em exactly} the entropy that arises in statistical mechanics, and this is the same as the entropy for these systems in the thermodynamic sense.

In thermodynamics, coming to thermal equilibrium means finding a state with maximal entropy given whatever constraints the system experiences.  In the present context, there are no heat flows and there is no notion of temperature or equilibrium. Instead \cite{jaynes_57},  the maximum entropy probability distribution provides us with a model that is consistent with observed facts---taking Eqs (\ref{match1}, \ref{match2}) as constraints---but otherwise has as little structure as possible.

To carry out the maximum entropy construction, we need to find the probability distribution that maximizes the entropy subject to the constraints in Eqs (\ref{match1}, \ref{match2}).  To do this we use the method of Lagrange multipliers \cite{multivar}.  We recall that if we want to maximize a function $f(\vec x )$ of many variables, $\vec x \equiv \{x_1 , x_2 , \cdots , x_D\}$ subject to the constraint that $g(\vec x ) = 0$, we can construct a new function $\tilde f (\vec x ; \zeta ) = f(\vec x ) + \zeta g( \vec x )$, where $\zeta$ is a ``Lagrange multiplier.'' If we maximize $\tilde f(\vec x ; \zeta)$ with respect to $\vec x$, we find a one parameter family of solutions, depending on the value of $\zeta$.  If we maximize again with respect to $\zeta$ we will pick out the one solution in this family that satisfies the constraint $g(\vec x ) = 0$.  If we have many constraints, we add more Lagrange multipliers, one for each constraint, and sum the corresponding contributions to $\tilde f$.

If we want to maximize the entropy of the probability distribution $P(\{\sigma_{\rm i}\})$ subject to the constraints in Eqs (\ref{match1}, \ref{match2}), the method of Lagrange multipliers tells us that we need to introduce a function
\begin{widetext}
\begin{eqnarray}
\tilde S [P(\{\sigma_{\rm j}\}); \{ h_{\rm i}, J_{\rm ij}\}] &\equiv& -\sum_{\seti} P(\seti) \ln P(\seti)  + \sum_{\rm i} h_{\rm i} \left[ \sum_{\{\sigma_{\rm j}\}}\sigma_{\rm i} P(\{\sigma_{\rm j}\}) - \langle\sigma_{\rm i}\rangle_{\rm expt}\right] \nonumber\\
&&\,\,\,\,\,\,\,\,\,\, + {1\over 2} \sum_{\rm ij} J_{\rm ij} \left[  \sum_{\{\sigma_{\rm k}\}}\sigma_{\rm i}\sigma_{\rm j} P(\{\sigma_{\rm k}\}) - \langle \sigma_{\rm i} \sigma_{\rm j}\rangle_{\rm expt}\right] + \lambda \left[ \sum_{\{\sigma_{\rm j}\}} P(\{\sigma_{\rm i}\})  -1\right] .
\end{eqnarray}
Here, $h_{\rm i}$ is the Lagrange multiplier introduced to enforce the constraint on $\langle \sigma_{\rm i}\rangle$ in Eq (\ref{match1}), and $J_{\rm ij}$ is the Lagrange multiplier introduced to enforce the constraint on $\langle\sigma_{\rm i}\sigma_{\rm j}\rangle$ in Eq (\ref{match2}); because the correlation matrix is symmetric we can take $J_{\rm ij} = J_{\rm ji}$, and the factor of $1/2$ reminds us that we are counting each term twice.  Finally, $\lambda$ is the Lagrange multiplier introduced to enforce the normalization of the probability distribution, which allows us formally to treat the variables $P(\{\sigma_{\rm i}\})$ as independent real numbers, not worrying that they have to sum to one.

To find the maximum of $\tilde S$, we take derivatives with respect to the elements of the probability distribution, and set these to zero:
\begin{eqnarray}
{{\partial \tilde S [P(\{\sigma_{\rm j}\}); \{ h_{\rm i}, J_{\rm ij}\}] }\over{\partial P(\{\sigma_{\rm i}\})}}
&=& -  \ln P(\{\sigma_{\rm i}\}) -1 + \sum_{\rm i} h_{\rm i} \sigma_{\rm i} + {1\over 2} \sum_{\rm ij} J_{\rm ij} \sigma_{\rm i}\sigma_{\rm j} + \lambda = 0\\
\Rightarrow  P(\{\sigma_{\rm i}\}) &=& {1\over{Z}} \exp\left[  \sum_{\rm i} h_{\rm i} \sigma_{\rm i} + {1\over 2} \sum_{\rm ij} J_{\rm ij} \sigma_{\rm i}\sigma_{\rm j} \right] ,
\end{eqnarray}
\end{widetext}
where $Z = e^{1-\lambda}$.  In addition, we need to maximize  $\tilde S$ with respect to the Lagrange multipliers.  For $\lambda$, the condition $\partial \tilde S [P(\{\sigma_{\rm j}\}); \{ h_{\rm i}, J_{\rm ij}\}] /\partial \lambda = 0$ is equivalent to the normalization condition, 
\begin{equation}
\sum_{\{\sigma_{\rm i}\}} P(\{\sigma_{\rm i}\})  =1.
\end{equation}
This sets the value of $Z$, which is called the partition function in statistical physics,
\begin{equation}
Z = \sum_{\{\sigma_{\rm i}\}} \exp\left[  \sum_{\rm i} h_{\rm i} \sigma_{\rm i} + {1\over 2} \sum_{\rm ij} J_{\rm ij} \sigma_{\rm i}\sigma_{\rm j} \right]  .
\end{equation}
If we maximize with respect to $h_{\rm i}$ we find the condition in Eq (\ref{match1}), and if we maximize with respect to $J_{\rm ij}$ we find the condition in Eq (\ref{match2}).  We have as many experimental measurements  as we have Lagrange multipliers, and so there are enough equations to determine all the parameters.  Solving these equations is another step,  discussed in the next Section.  Strictly speaking, finding the point where derivatives vanish yields an extremum, not necessarily a maximum of the entropy.  But the entropy is a convex function of the probability distribution \cite{cover+thomas_91}, so that  relevant second derivatives all are negative; hence any extremum will be a maximum.

To summarize, we have shown that the least structured probability distribution consistent with measured averages and pairwise correlations has the form
\begin{eqnarray}
P(\{\sigma_{\rm i}\}) &=& {1\over Z} e^{-E(\{\sigma_{\rm i}\})}\label{boltzA1}\\
Z &=& \sum_{\{\sigma_{\rm i}\}} e^{-E(\{\sigma_{\rm i}\})}\label{boltzA2}\\
E(\{\sigma_{\rm i}\}) &=& -\sum_{\rm i} h_{\rm i} \sigma_{\rm i} - {1\over 2} \sum_{\rm ij} J_{\rm ij} \sigma_{\rm i}\sigma_{\rm j} .\label{boltzA3}
\end{eqnarray}
The parameters $\{h_{\rm i} , J_{\rm ij}\}$ are not arbitrary, but must be adjusted to be sure that the predicted averages and pairwise correlations match the measured values, as in Eqs (\ref{match1}, \ref{match2}).  Equations (\ref{boltzA1}, \ref{boltzA2}) are exactly the Boltzmann distribution, which describes the distribution of  states taken on by a system in thermal equilibrium, where the energy of each state is given by $E(\{\sigma_{\rm i}\})$.  In the physical setting, there is a real temperature $T$, which determines an energy scale $k_B T$, where $k_B$ is Boltzmann's constant.   Then, to be precise, we should write
\begin{equation}
P(\{\sigma_{\rm i}\}) = {1\over Z} e^{-E(\{\sigma_{\rm i}\})/k_B T} ,
\end{equation}
but we are free to choose our units of energy so that $k_B T = 1$.  Then it is clear that our problem of building minimally structured models leads us exactly to a statistical mechanics model of the system we are studying.

The physical interpretation of our model is that the (yes/no) votes of judges are Ising ($+1/-1$) spins that each experience a ``magnetic field'' $h_{\rm i}$ and interact in pairs through the couplings $J_{\rm ij}$.  We have chosen a sign convention such that $h_{\rm i}>0$ favors a yes vote ($\sigma_{\rm i}=+1$), and $J_{\rm ij} >0$ favors justices $\rm i$ and $\rm j$ voting in the same way.

Because we are asking to match both the averages $\langle\sigma_{\rm i}\rangle$ and the pairwise correlations $\langle\sigma_{\rm i}\sigma_{\rm j}\rangle$,  there are two sets of terms in the ``energy'' $E(\{\sigma_{\rm i}\})$.  In our initial formulation of the voting problem on the US Supreme Court, as described in the main text, yes and no votes are symmetric, and we automatically have $\langle\sigma_{\rm i}\rangle =0$ for every justice $\rm i$.  Then we need to match only the correlations between the votes of pairs of justices, and hence the energy function simplifies to
\begin{equation}
E(\{\sigma_{\rm i}\}) =  - {1\over 2} \sum_{\rm ij} J_{\rm ij} \sigma_{\rm i}\sigma_{\rm j} .\label{boltzA4}
\end{equation}
We see that Eqs (\ref{boltzA1},  \ref{boltzA4}) are the same as Eqs (\ref{ising1}, \ref{ising2}) of the main text.  In a later discussion, we will break the symmetry between yes and no votes, and the fields $h_{\rm i}$ will then be important (Appendix \ref{sec:ideological_votes}).

It is important to emphasize that the maximum entropy method is {\em not} a model.  It is a framework for building models that capture particular aspects of the data while making no additional assumptions.  Thus, there are no free parameters to be ``fit,'' and we are able to make unambiguous, quantitative predictions, as in Fig \ref{fit}.

\section{Solving the inverse problem}\label{sec: solving inverse}
The maximum entropy construction arrives at Eqs (\ref{boltzA1}, \ref{boltzA2}, \ref{boltzA4}) analytically.  To complete the construction, we actually have to find the numerical values of the coupling parameters $J_{\rm ij}$ that allow the model to match the observed correlations.  That is, we have to solve Eq (\ref{match2}), which can be written more explicitly as
\begin{equation}
\sum_{\{\sigma_{\rm k}\}} \sigma_{\rm a}\sigma_{\rm b} {1\over Z} \exp\left[ {1\over 2}\sum_{\rm ij} \sigma_{\rm i}J_{\rm ij}  \sigma_{\rm j} \right] = \langle\sigma_{\rm a}\sigma_{\rm b}\rangle_{\rm expt} .
\end{equation}
We note that both the correlations and the couplings define symmetric matrices.  Thus, with ${\rm i} = 1,\, 2,\,\cdots ,\, N =9$ justices, these are $N(N-1)/2=36$ simultaneous equations for the $36$ independent parameters $J_{\rm ij}$.  These equations are relatively straightforward to solve numerically, for example using MATLAB's {\tt fsolve} routine.  Results for the second Rehnquist court are shown in Fig \ref{Jij}.

\begin{figure}\centering
  \includegraphics[width=\linewidth]{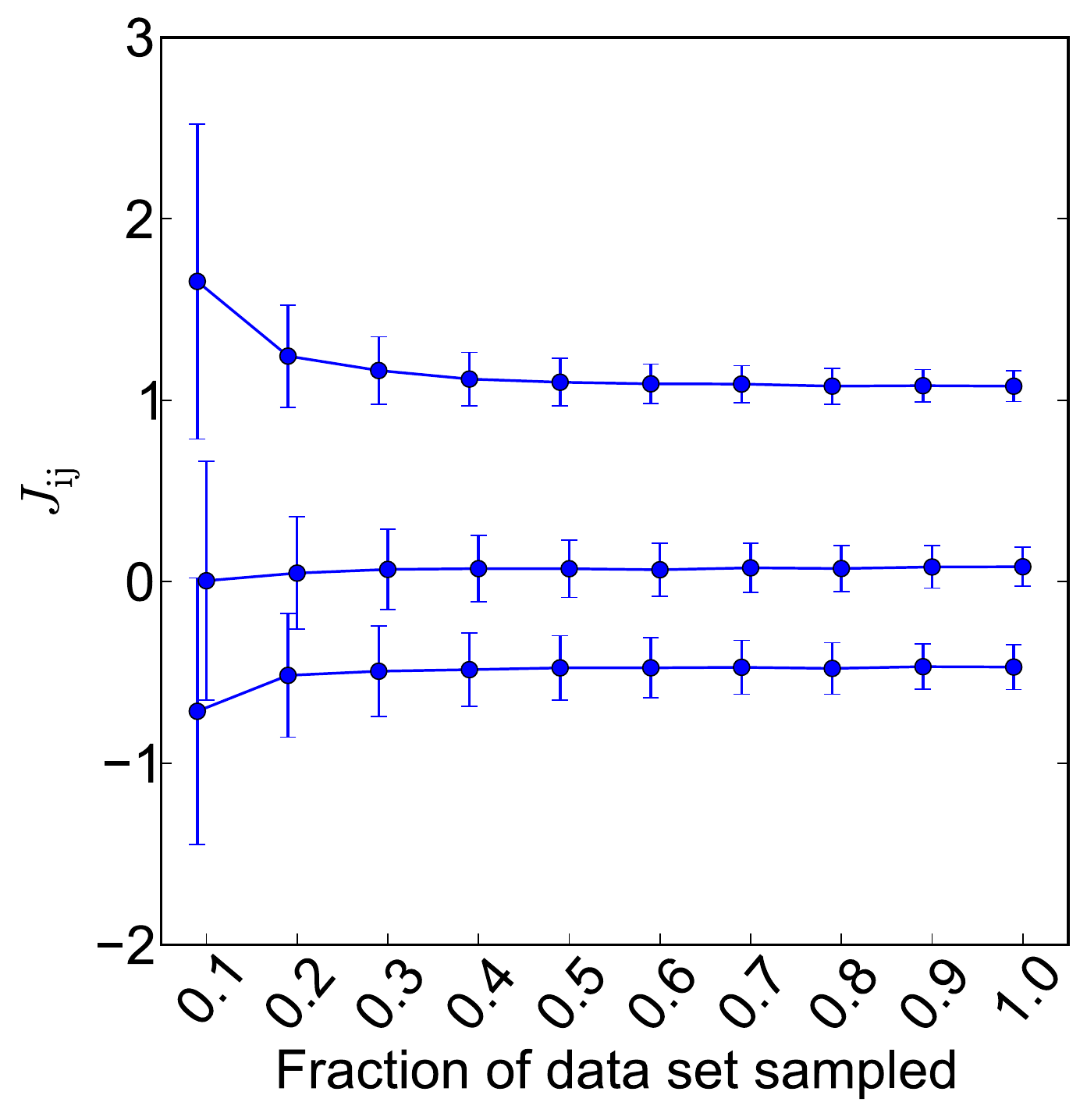}
  \caption{Convergence of parameters with larger sample sizes.   As examples, we show the positive and negative couplings with largest absolute values, as well as the coupling that has the smallest absolute value.  We take bootstrap samples with replacement, of size $n=f\times N$ with $N=895$ the total number of votes recorded for the  second Rehnquist Court.  We construct independent maximum entropy models for each sample, and plot the mean couplings vs. sample size $f$;  error bars are standard deviations of the couplings over multiple bootstrap samples at each $f$.  Points have been slightly displaced along the x-axis to minimize overlap between error bars.
  \label{graph:convergence}}
\end{figure}

\begin{figure}\centering
  \includegraphics[width=.8\linewidth]{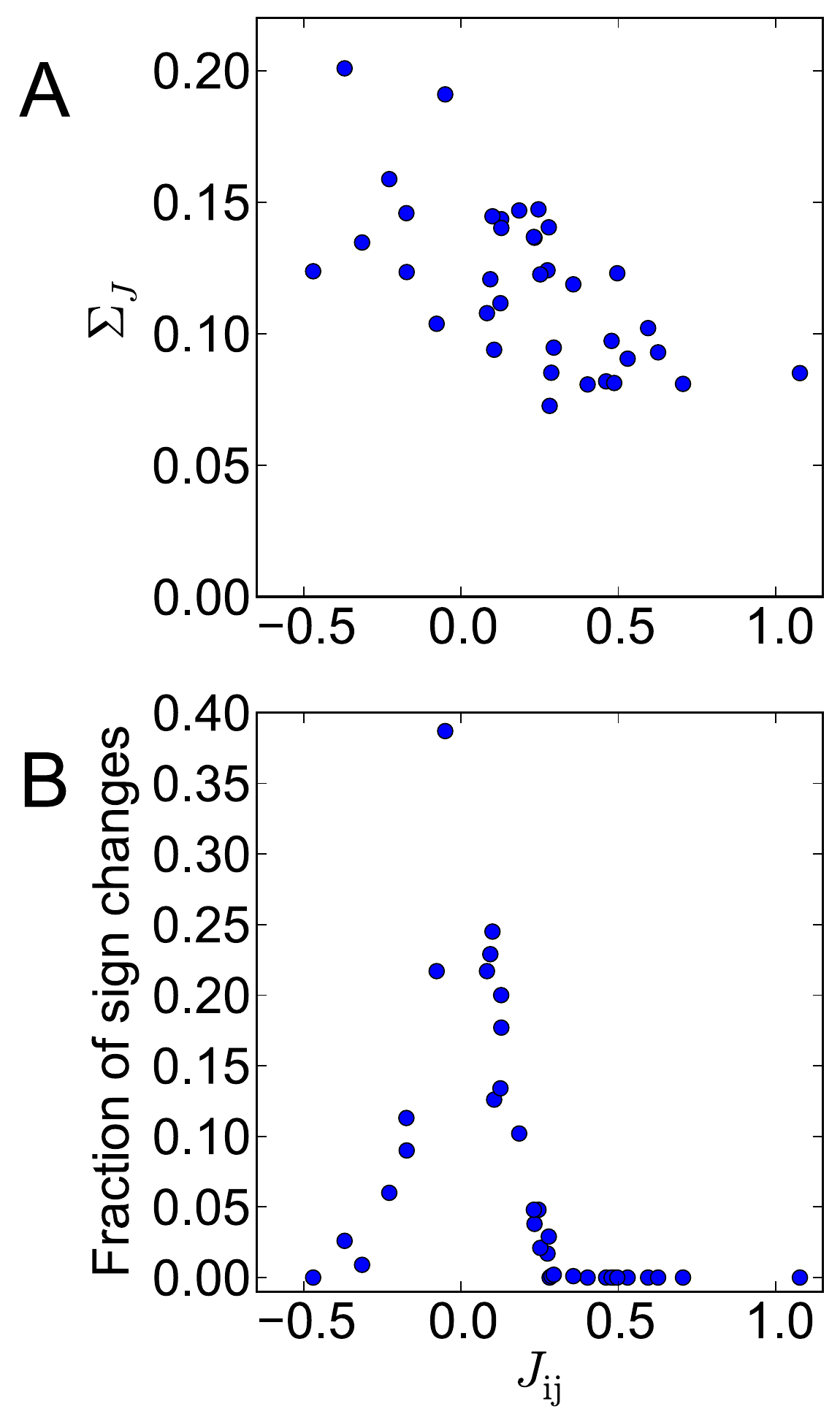} 
  \caption{(A) Standard deviation $\Sigma_J$ vs. $J_\mr{ij}$. (B) Probability of a sign flip in $J_\mr{ij}$ over bootstrap samples.  The strongest 2/3 of couplings have fixed sign with $95\%$ confidence. \label{graph:Jijsigns}}
\end{figure}

The maximum entropy problem is formulated on the assumption that we know the correlation functions from experiment.  In fact, these measurements come with error bars, since our sample is finite.  We would like to convince ourselves that these errors have only a small impact on our ability to construct the model and make predictions.  As a first test, we ask what happens when we choose random fractions of the full data set.  Figure \ref{graph:convergence} shows that, for selected elements of the matrix $J_{\rm ij}$, our best estimate has a very weak systematic dependence on the size of the data set, and that these individual parameters can be determined with reasonable precision.  Fig \ref{graph:Jijsigns} surveys these errors in the entire matrix, showing that all elements are determined within $\pm 0.2$, and many within $\pm 0.1$; these random errors are estimated, as in Fig \ref{graph:convergence}, across bootstrap resamplings of the data.  Importantly, for most of the $J_{\rm ij}$, these resamplings have a low probability of changing our estimate of the sign of the interaction, and uncertainty about the sign is confined to the $J_{\rm ij}$ that have the smallest magnitude.

While our model is parameterized by the $J_{\rm ij}$, the fundamental prediction of the model is the probability of each voting pattern; the logarithm of this probability is the ``energy'' of the state, from Eq (\ref{ising2}).  If the errors in all the $J_{\rm ij}$ were independent, then the errors in the energy would typically be six times larger than the errors in the individual $J_{\rm ij}$, and this would be quite bad.   In fact the errors are unlikely to be independent, and they are not.  To get some intuition, we know that if the $J_{\rm ij}$ themselves are drawn at random, then the correlations $C_{\rm ij}$ have a complicated structure \cite{mezard+al_87}.  Conversely, we expect that independent random additions to the $C_{\rm ij}$ would produce a structured change in the $J_{\rm ij}$.

\begin{figure}[b]\centering
  \includegraphics[width=.8\linewidth]{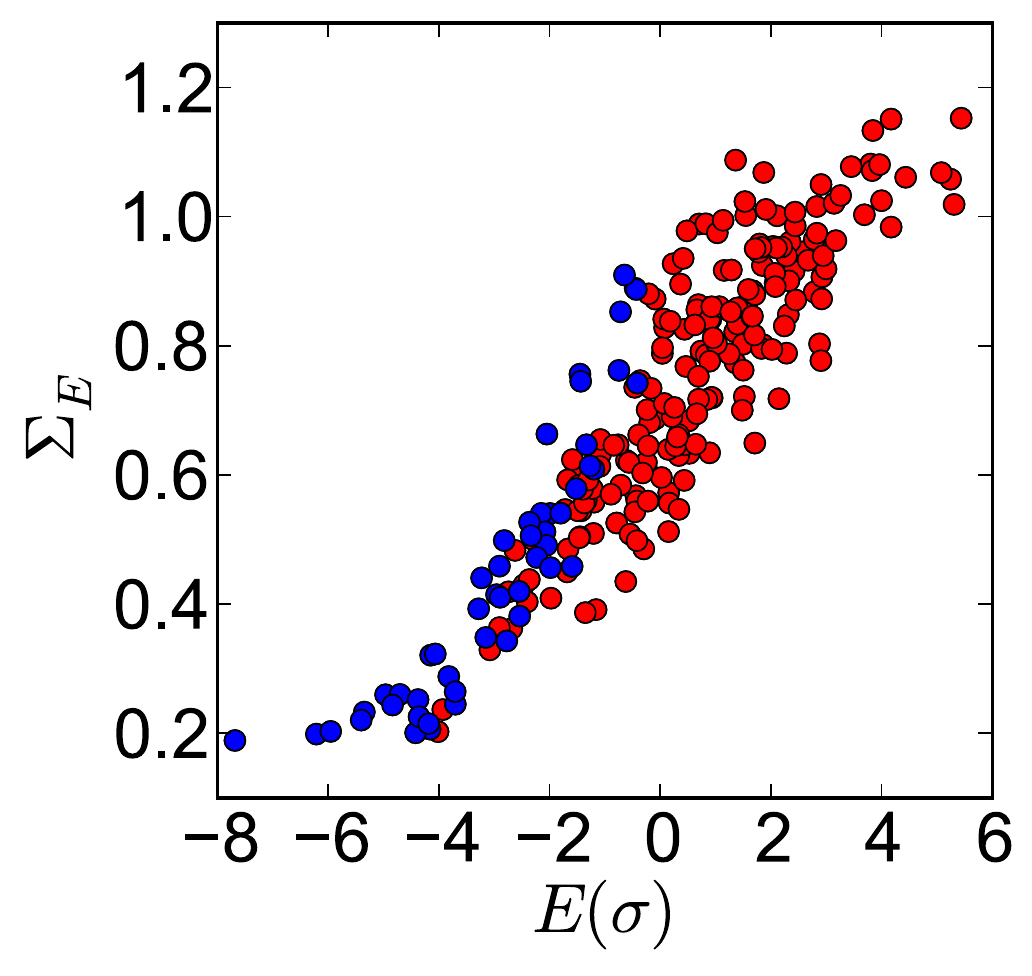}
  \caption{Standard deviation $\Sigma_E$ vs. mean of energies $E(\s)$, for each state observed in the data.   States that appear only once are shown in red, and states that appear more than once are shown in blue.   \label{E_errors}}
\end{figure}

When we draw random samples of the data to generate $C_{\rm ij}$ and construct the corresponding $J_{\rm ij}$, we get the whole matrix of $J_{\rm ij}$ and hence a whole set of predictions for the energies of individual states.  We can look at the standard deviations of these energies, as a function of the means, as shown in Fig \ref{E_errors}.  Low energy (more likely) states have errors $\Sigma_E \sim 0.2$, which means that we can predict the probability of these states with $\sim 20\%$ accuracy.  This is possible only because the errors in the $J_{\rm ij}$ are correlated.  Once we reach $\Sigma_E \sim \ln 2$, we can predict probabilities only within a factor of two.  But this level of error is reached only for states with energy $E$ roughly $8$ units above the lowest energy state, and hence relative probability $\sim e^{-8}< 10^{-3}$.  With only $895$ samples, we thus should be able to predict, with reasonable reliability, the probabilities of all voting combinations that actually occur in the data, and this is borne out in Fig \ref{fit}.

\section{The energy landscape}\label{app:landscape}

Maximum entropy models express the probability of a system being in a certain state (here, the pattern of votes by the nine justices) in terms of an ``energy'' for that state, and this energy in turn is built out of terms that express ``interactions'' among the elements of the system (here, the votes of the individual justices).    It is useful to think about the energy as a landscape on the space of states, with deep valleys corresponding to states of high probability and mountains corresponding to states of low probability; mountain passes provided the most likely paths that connect one highly likely valley to another.   The model defines the energy for all possible states, not only those which are observed to occur in our finite sample of data.  Further, in the approach we have taken here, the entire landscape is determined by the {\em measured} correlations among the votes of pairs of justices.  Thus, the landscape here is not a metaphor, but something that we can construct explicitly and quantitatively, with no free parameters.

\begin{figure*}[t]\centering
 \includegraphics[width=\linewidth]{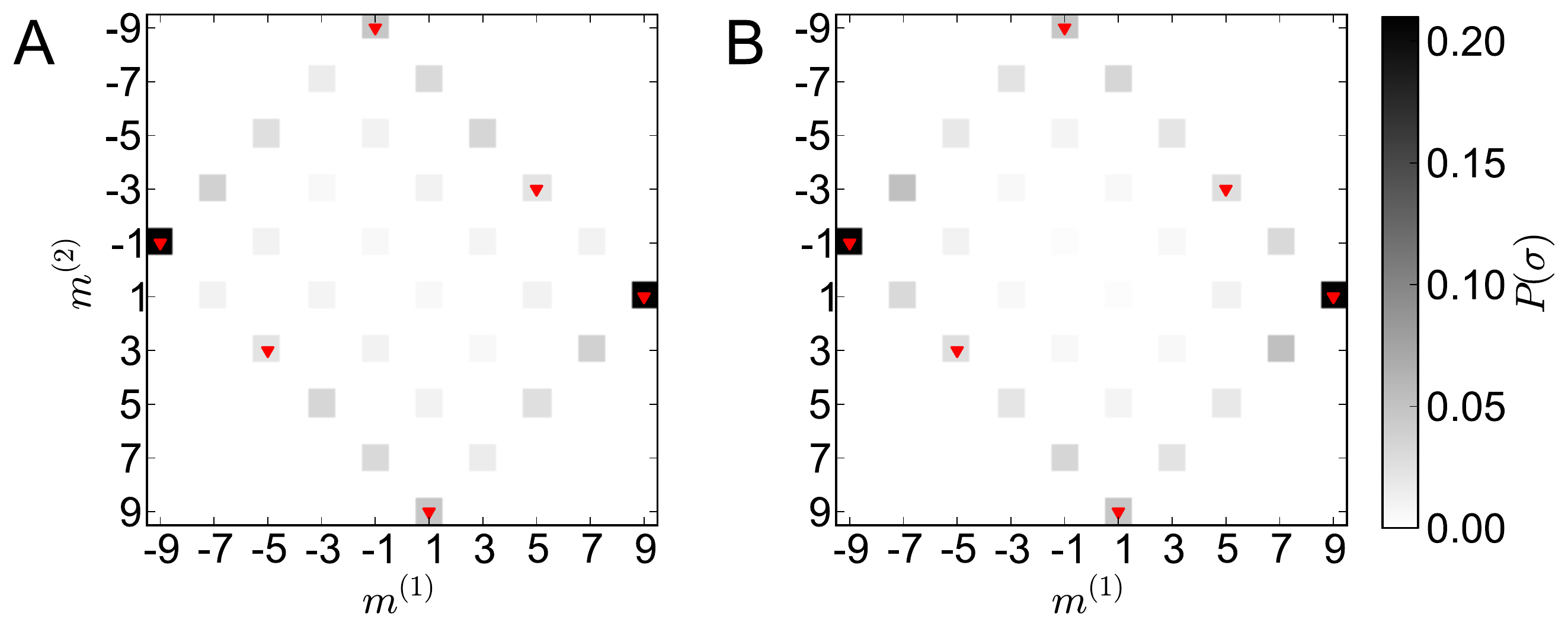}
  \caption{Projection of the energy landscape in the data (A) and the predictions of the pairwise maximum entropy model (B). The horizontal axis shows the projection $m^{(1)}$ onto the unanimous $+1$ basin, and the vertical axis shows the projection $m^{(2)}$ onto 5--4 basin oriented so the majority voters are $+1$. The 7--2 basin lies in between as expected. Local energy minima are marked with red triangles. Note that points are separated by at least one empty block because a single vote flip corresponds to a change in 2 along either dimension. The space is highly structured, with density almost exclusively on the periphery and with a nearly empty center.}\label{E_proj}
\end{figure*}

In the case of the Supreme Court, the space of states is discrete:  each justice votes yes or no ($\sigma_{\rm i} = +1$ or $\sigma_{\rm i} = -1$), and so while there are nine dimensions the allowed states live only on the corners of a (hyper)cube in these nine dimensions.  Nonetheless we can identify two states as being neighbors if the jump from one state to the other involves changing the vote of only one justice.  The bottom of a valley is a place where all moves take us uphill, and correspondingly we can ask for local minima of the energy function such that flipping the vote of any one justice always increases the energy.  These local minima of the energy are predicted to be local maxima of the probability, and hence provide us with anchor points for thinking about the voting patterns.  A local minimum of energy defines a prototypical vote, and the neighboring patterns---which are predicted to be less likely---can be thought of as ``noisy versions'' of this prototype.

As discussed in the main text, the model that we find for the patterns of votes on the US Supreme Court belongs to a class of models known in the physics literature as spin glasses \cite{mezard+al_87}, and a signature of these models is that their energy landscapes have multiple local minima.  Thus, we expect that even our small ($N=9$) system will have several local minima or prototypical voting patterns, and this is what we find.

Concretely, the energy landscape for the second Rehnquist court has three major valleys, plus the symmetric mirror of these valleys obtained by exchanging the definitions of yes and no.   Together, the states in these valleys account for over 99\% (508/512) of the possible voting states, corresponding to almost the full mass of the probability distribution.  As noted in the main text, the prototypical states at the bottoms of the valleys are the unanimous vote, the 5--4 ideological split, and the 7--2 vote against Scalia and Thomas.   We emphasize that these breaks along ideological lines emerge from the model  even though we make no reference to ideology in our construction; these structures are encoded in the pairwise correlations, and the maximum entropy method allows us to make these structures explicit.

Leaning on the equivalence to statistical physics, we can think of the local minima in energy as the states into which the system is ``trying'' to order.   If $\{\xi_{\rm i}^{(n)}\}$ is the $n^{\rm th}$ local minimum, we can measure how close the system has come to this state by the overlap
\begin{equation}
m^{(n)} = \sum_{{\rm i}=1}^N \xi_{\rm i}^{(n)}\sigma_{\rm i} .
\end{equation}
If the system is very deep in the valley defined by $\{\xi_{\rm i}^{(n)}\}$, then we will have $m^{(n)}\approx N$.   With two dominant valleys in the energy landscape, the unanimous vote and the 5--4 ideological split, there are two natural ``order parameters'' $m^{(1)}$ and $m^{(2)}$, respectively.  In Fig \ref{E_proj}, we show the probability distribution projected onto these two dimensions, both for the real data and as predicted by the maximum entropy model.

In the projections along ($m^{(1)}$, $m^{(2)}$), we can see the clear local maxima of probability when $m = \pm N$, both in the data and in the predictions of the model.    More surprising is that the distribution is almost confined to the edge of the allowed space.  This feature of the data, which is predicted clearly by the model, means that the full distribution is in effect dominated by the competition between the tendencies toward unanimity and ideological division, and this is not just a qualitative statement but a quantitative one.  Importantly, all of this structure is predicted by the maximum entropy model using only the observed pairwise correlations among votes as inputs.

The idea of projecting the pattern votes onto two dimensions is not new.  Ten years ago, Sirovich \cite{sirovich2003} noted that the covariance of justices' votes is dominated by two principal components, which are very close to the dimensions defined by $m^{(1)}$ and $m^{(2)}$ here.    Although we start with the same covariance matrix, the maximum entropy approach does more than identify dominant dimensions, since it makes quantitative predictions about the entire probability distribution.    This is possible because the models we build respect the discrete nature of the votes---we are constructing a joint probability distribution for binary variables---while the covariance matrix itself could have arisen from a set of continuous variables, and the geometric interpretation of principal components analysis does not make reference to the ``corners of the hypercube'' structure in the space of voting patterns.  By respecting the discreteness of votes, even the least structured model that is consistent with the covariance matrix exhibits a very rich structure, and one that is in detailed agreement with the data.

\section{Estimating entropies and information}\label{app:ent_est}
At various points in our analyses, we estimate information theoretic quantities such as the entropy of a distribution or the mutual information between different variables.  It is well known that such estimates can be systematically biased in small data sets \cite{miller_55}.  This problem received considerable attention in the analysis of experiments on neural coding \cite{treves+panzeri_95, strong+al_98, paninski_03, nemenman+al_04}, and here we explain how we use what was learned in that context to be sure that our estimates are reliable.  For a more pedagogical discussion, see Appendix A.8 of Ref \cite{bialek_12}.

For the second Rehnquist Court, it is plausible that all relevant information theoretic quantities  will be well determined:  there are  $\Omega = 512$ states, or really $256$ independent probabilities in the symmetrized data, and we have $N = 2\times 895$ samples (again, doubled because of symmetry).    For other natural courts (cf Fig \ref{consensus}), however, the number of samples is highly variable,  down to as few as 91 votes.  We would like to use all of the available examples, and thus we need to understand how the limited data set sizes can bias our estimates.

\begin{figure}[tb]\centering
  \includegraphics[width=.8\linewidth]{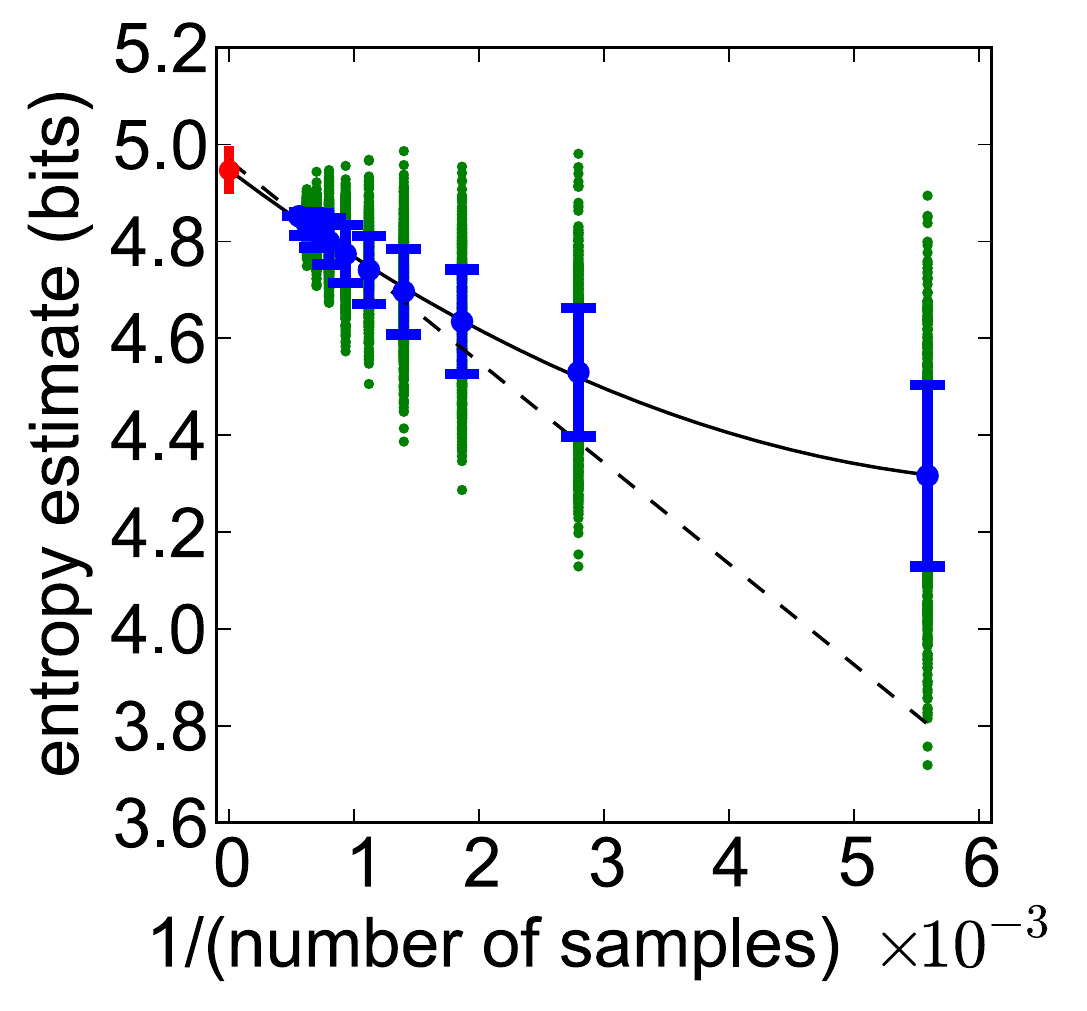}
  \caption{Estimate of entropy voting patterns in the second Rehnquist Court. Given an initial data set with $N$ samples, we draw multiple bootstrap samples of size $n$ and form the ``naive'' estimate of the entropy from Eq (\ref{eq:Snaive}); results are shown as green points; means and standard deviations at each $n$ plotted in blue.  Extrapolations based on Eq (\ref{eq:S_expansion}): linear fit to samples $n>N/2$ is the dashed line, and the quadratic fit to all plotted points is shown as a solid line. Red point is our best estimate, with errors. \label{S_est}}
\end{figure}

The hardest quantity to estimate is the entropy of the distribution of voting patterns, since this depends on the probability of every single state.  The na\"{i}ve approach is to identify the observed frequency of occurrence of each state with its probability, and then plug these estimates into the definition of the entropy, here measured in bits,
\begin{equation}
S_{\rm naive}(n) = -\sum_{\{\sigma_{\rm i}\}} {\hat P}_n (\{\sigma_{\rm i}\}) \log_2 {\hat P}_n (\{\sigma_{\rm i}\}) ,
\label{eq:Snaive}
\end{equation}
where ${\hat P}_n$ is the frequentist estimate of probability based on $n$ samples.  The differences between frequencies and probabilities are random---they average to zero, and as the sample size becomes larger their variance decreases uniformly.  But the entropy is a nonlinear function of the probabilities, and so these random errors become systematic \cite{miller_55,treves+panzeri_95}.  If the number of samples $n$ is large enough, these systematic errors in the naive estimate take a simple form,
\begin{equation}
S_{\rm naive}(n) = S_\infty + {A\over n} + {B\over{n^2}} + \cdots, \label{eq:S_expansion}
\end{equation}
where $S_{\infty}$ is the true entropy that we would find with an infinite number of samples, and $A$ and $B$ are constants.  If we can convince ourselves that we are in the regime where this formula describes our systematic errors, then we are safe in taking the extrapolated $S_\infty$ as an estimate of the entropy, as shown in Fig \ref{S_est}.  Notice that the ``finite size correction,'' $S_{\infty} - S_{\rm naive}(N)$ in Fig \ref{S_est}, is less than $10\%$ of the total entropy, and that the difference between extrapolations where we include or ignore $B$ in Eq (\ref{eq:S_expansion}) is even smaller; this is true, consistently, for the data on all the natural courts that we consider.

The extrapolation procedure in Fig \ref{S_est} should generate an unbiased estimate of the actual entropy.  The maximum entropy models that we have constructed should, by definition, generate an upper bound on the entropy.  This upper bound is based on measurements of the pairwise correlations, and since there are only $36$ independent correlation matrix elements, even small data sets give a fairly reliable basis from which to construct these models.  Happily, there is also a lower bound on the entropy that we can construct, and this too is rather robust to small sample sizes.

In a uniform distribution over $\Omega$ states, the probability that two states chosen at random are the same is $P_c = 1/\Omega$ and the entropy is $S= \log_2\Omega$.  Thus, in this case, we can estimate the entropy if we can estimate the probability of a coincidence, where two states are the same. Notice that if we have $n$ samples, we can test $n(n-1)/2$ independent pairs, and so we start to get a reliable estimate of $P_c$ as soon as $n \gg \sqrt{\Omega}$, which is much less than the naive expectation that we need to see all the states ($n\sim \Omega$) in order to say something about the distribution from which they are drawn.  As an aside, this is the basis for the ``birthday problem,'' where the number of people needed to make it likely that two of them share the same birthday is much less than the number of possible birthdays; a discussion appears in Feller's classic text \cite{feller}.

To go beyond the uniform distribution, we note that the probability of a coincidence among two randomly chosen states is
\begin{equation}
P_c = \sum_{\{\sigma_{\rm i}\}} \left[ P(\{\sigma_{\rm i}\})\right]^2  = \langle P(\{\sigma_{\rm i}\})\rangle ,
\label{defPc}
\end{equation}
where $\langle \cdots \rangle$ stands for an expectation value over the distribution $P(\{\sigma_{\rm i}\})$.  But for any positive random variable $x$, we have
\begin{equation}
\log_2 \langle x \rangle \geq \langle \log_2 x\rangle .
\end{equation}
Applying this inequality to Eq (\ref{defPc}), we have
\begin{eqnarray}
\log_2 P_c = \log_2 \langle P(\{\sigma_{\rm i}\})\rangle &\geq& \langle \log_2 P(\{\sigma_{\rm i}\})\rangle ,\\
\Rightarrow -\log_2 P_c &\leq& -  \langle \log_2 P(\{\sigma_{\rm i}\})\rangle .
\end{eqnarray}
But
\begin{equation}
-  \langle \log_2 P(\{\sigma_{\rm i}\})\rangle  = -\sum P(\{\sigma_{\rm i}\})\log_2 P(\{\sigma_{\rm i}\}) = S,
\end{equation}
and so we have a lower bound on the entropy,
\begin{equation}
S \geq S_{\rm Ma} = -\log_2 P_c .
\end{equation}
We will refer to this as the ``Ma bound,''  after Ref \cite{Ma}.

\begin{figure}[tb]\centering
  \includegraphics[width=.8\linewidth]{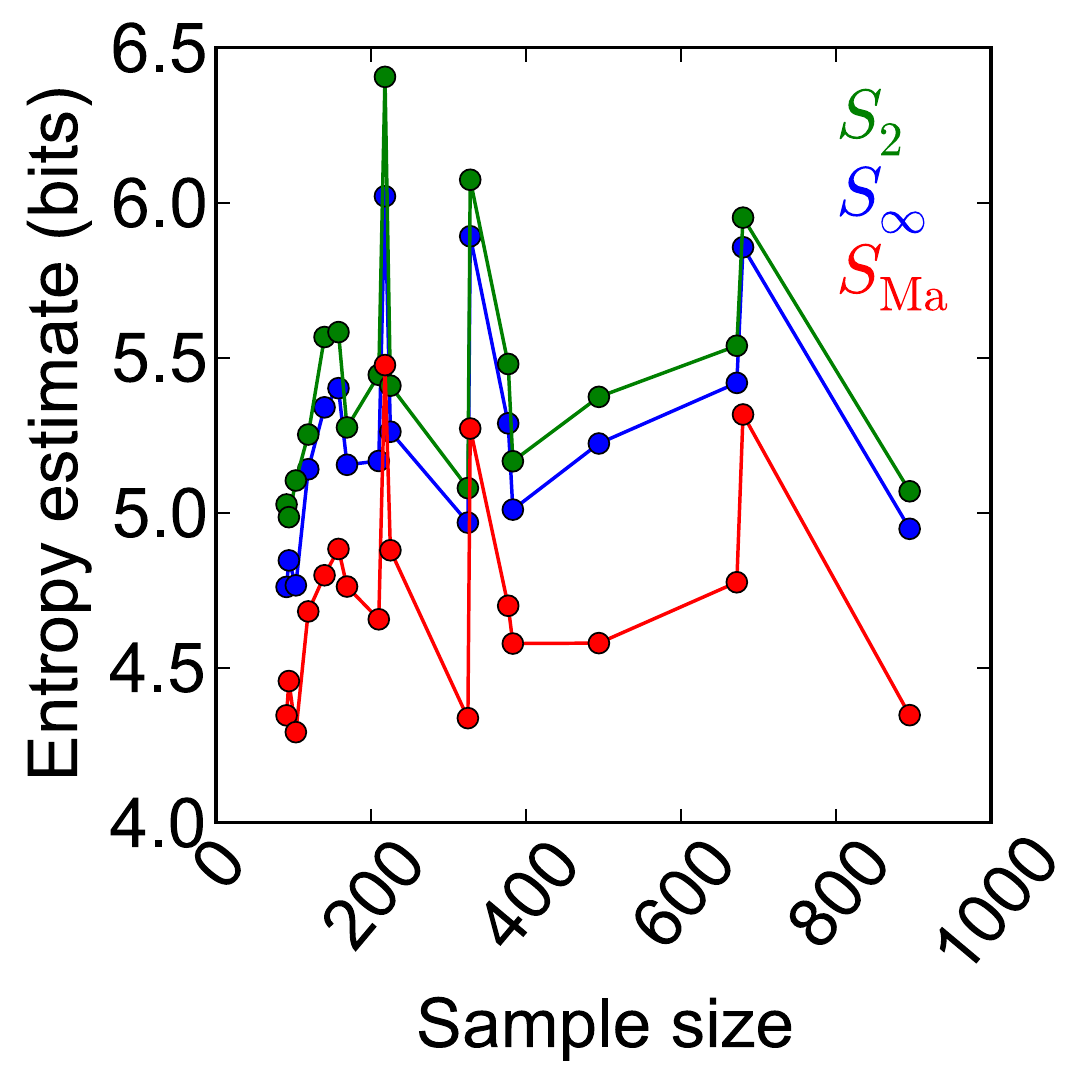}
  \caption{Entropy estimates for all the natural courts, as a function of sample size. Statistical errors, as shown in Fig \ref{S_est}, are small.  \label{bounds_fig}}
\end{figure}

In Figure \ref{bounds_fig} we show the various entropy estimates for all the natural courts we consider, ordered by the number of votes recorded for each court (sample size).  We see that entropy estimates based on extrapolation, as in Fig \ref{S_est}, are consistently $\sim 10\%$ above the Ma bound, and this is true across the full range of sample sizes.  Indeed, the entropies (and the Ma Bounds) for the different natural courts themselves vary by only $\pm 10\%$, suggesting that the structure of voting patterns is quite stable across the decades.  Although there are $2^9 = 512$ possible voting patterns, the fact that the entropy is consistently $S \sim 5\,{\rm bits}$ indicates that, effectively, the court uses only $2^S \sim 32$ of these patterns.  But with these few patterns, even one hundred votes is enough to generate a reasonably good sampling.   It should be noted that entropy estimates based on extrapolation can easily violate the Ma bound if the number of samples we have in the data is genuinely too small (see, for example, Ref \cite{strong+al_98}).  Taken together, these results suggest strongly that our entropy estimates are reliable for all of the natural courts, and hence we can compute the multi--information and assess the fraction of this which is captured by the maximum entropy model, as discussed in the text.

For the symmetrized data, each justice is equally likely to vote yes or no, and hence if they voted independently the entropy would be exactly $S_1 = 9\,{\rm bits}$.  Then we can read from Fig \ref{bounds_fig} our estimate of the multi--information, $I_K^{\rm est} = S_1 - S_\infty$, as well as the multi--information captured by the pairwise maximum entropy model, $S_1 - S_2$.  The resulting fraction $F = (S_1 - S_2)/I_K^{\rm est} = 0.95\pm 0.03$, where the error bar is the standard deviation across the set of natural courts.  Alternatively, we know that the multi--information must be greater than $I_K^{\rm Ma} = S_1 - S_{\rm Ma}$, and we thus can conclude that $F \geq (S_1 - S_2)/I_K^{\rm Ma} = 0.83 \pm 0.03$; this is a true bound, and hence a conservative estimate.

We also need to estimate other information theoretic quantities, such as the mutual information between individual justices' votes and the court majority, $I(\sigma_{\rm i}; \gamma )$. But these quantities involve probability distributions over many fewer states, and thus the sampling issues discussed here are negligible.

\begin{figure*}[thb]\centering
  \includegraphics[width=.8\linewidth]{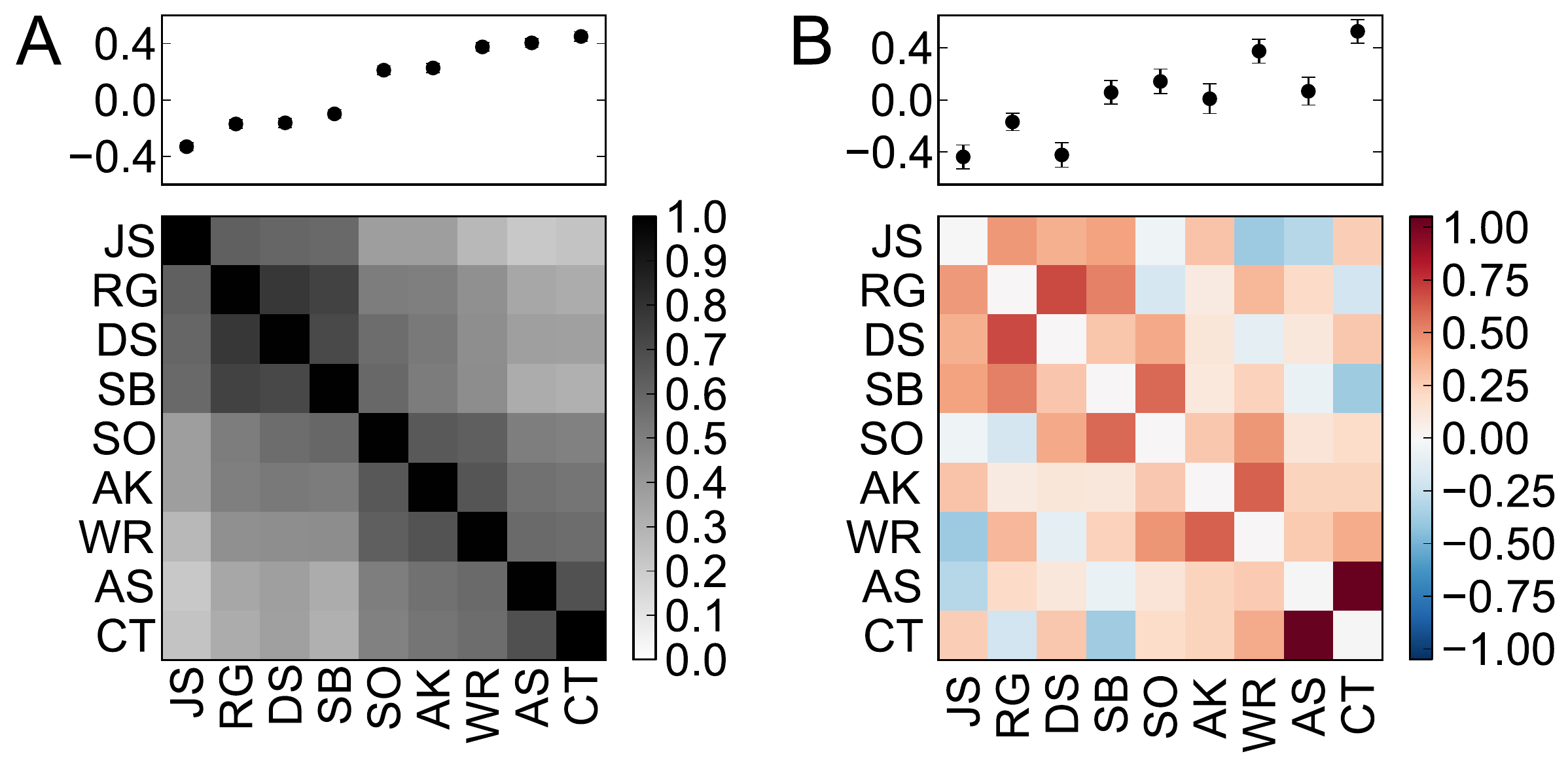}
  \caption{Data and model  for ideologically labeled votes on the second Rehnquist court. (A) Correlation matrix $ \langle \sigma_{\rm i}\sigma_{\rm j} \rangle$ of ideological votes in the Rehnquist court, with average votes $\br{\s_\mr{i}}$ plotted above. As explained in the text, conservative/liberal votes are represented as binary variables $\sigma_{\rm i} = \pm1$ for each justice $\rm i$ \cite{data}.    (B) Effective interactions  $J_\mr{ij}$, with the biases $h_\mr{i}$ plotted above.  \label{graph: Cij, Jij ide}}
\end{figure*}

\section{Ideologically defined votes\label{sec:ideological_votes}}

As noted in the main text, the definition of ``yes'' and ``no'' votes in Supreme Court decisions has an element of arbitrariness, since the question before the Court is always to affirm or overturn a previous decision.  In our initial approach to the data, we elevated this arbitrariness to a symmetry, imagining that every case could have come with the opposite definition of yes and no, so that voting patterns $\{\sigma_{\rm i}\}$ and $\{-\sigma_{\rm i}\}$ should be equally probable.  An alternative is to note that each case presents an issue that can be mapped to the state of national politics, and there is (except for rare cases) a reasonable consensus that someone with leftist tendencies would vote one way and someone with rightist tendencies the other.  This is, of course, only one dimension along which cases may vary.

So much attention is paid to the right/left split in today's politics that we might imagine such influences are dominant \cite{Segal2002}. Indeed, it could be  that each justice  responds independently to the merits of each case as seen through his or her political biases, and that what we see as correlations reflects nothing more than the fact that we are averaging over cases with different features.   It may be useful to make the analogy to the case of sensory neurons.  If each neuron in a network responds independently to its sensory inputs, and we average over these inputs, we will see correlations among the responses of different neurons. If we can hold the inputs fixed, however, it is possible that the correlations will vanish because there are no genuine interactions among the cells.   It is not clear that we can do a completely analogous experiment with the Supreme Court, but mapping each yes/no vote onto a right/left decision seems like a reasonable start, as emphasized previously \cite{Martin2004,sirovich2003,Kemp2008}.

In fact, the raw data of Ref \cite{data} come labeled by the right/left sign of each vote, although we note some difficulties, emphasized also in Ref \cite{data}.  First, there is a problem of circularity, since  ideologies are defined partly by the actors themselves. Thus, it is not truly an external, fixed measure along which we can consider the votes of SCOTUS. Instead, the axis is partially defined by the internal dynamics of the system. The problem of circularity then implies a second problem of non-stationarity, since the definition of liberal and conservative positions evolve over time. Spaeth et al. have adjusted for these changes over time, but it is   difficult  to gauge what the association between ideological ideals and votes may be at a given time. Overall, there is evidence for an important unidimensional space similar to our intuitive concepts of conservative vs. liberals, but how this axis overlaps with our intuitions seems inexact.

If we take the suggestion of ideological bias seriously, there are established definitions for what constitute the two ends of the ideological axis. Spaeth et al. have used these definitions to classify the votes as liberal (which we assign as $\sigma_{\rm i} =-1$) or conservative ($\sigma_{\rm i} = +1$) \cite{data}. The votes that do not fall along this classification have been removed from both symmetrized and ideological analyses (roughly 2\% of votes in the second Rehnquist Court). Figure \ref{graph: Cij, Jij ide} shows the mean votes and pairwise correlations $ \br{\s_\mr{i}\s_\mr{j}}$ in the ideologically calibrated data.

\begin{figure}[b] 
\centerline{  \includegraphics[width=.8\linewidth]{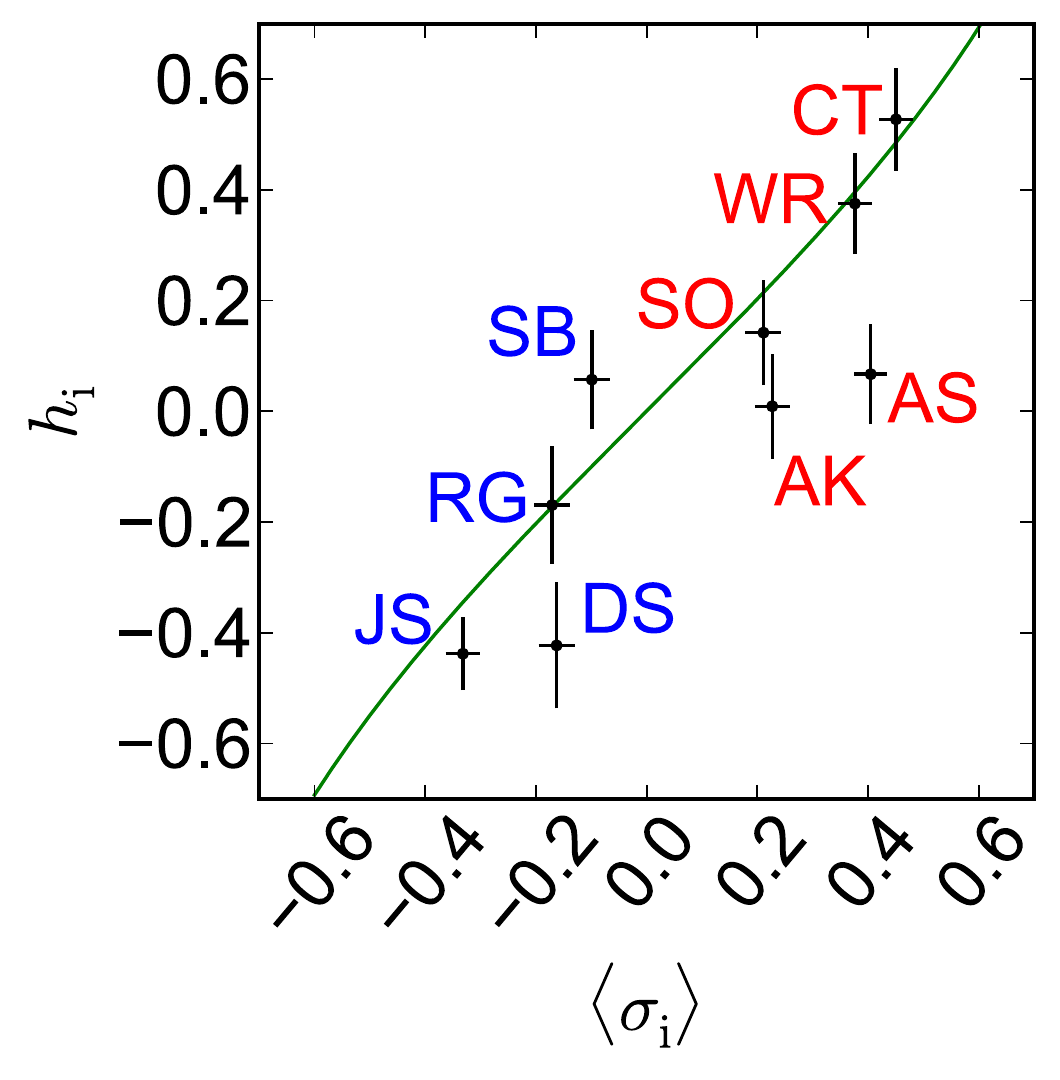}}
  \caption{Bias $h_\mr{i}$ against mean votes $\br{\s_\mr{i}}$. Green line traces the function $\langle\sigma\rangle = \tanh(h)$, as would be expected if each justice voted independently with bias $h_{\rm i}$. Error bars on $h_{\rm i}$ are the standard deviation across independent constructions of maximum entropy models from multiple bootstrap samples, as for $\Sigma_J$ in Fig \ref{graph:Jijsigns}, while errors in $\langle\sigma_{\rm i}\rangle$ arise as usual from counting statistics. \label{h_vs_s}}
\end{figure}

\begin{figure}[t]\centering
  \includegraphics[width=.8\linewidth]{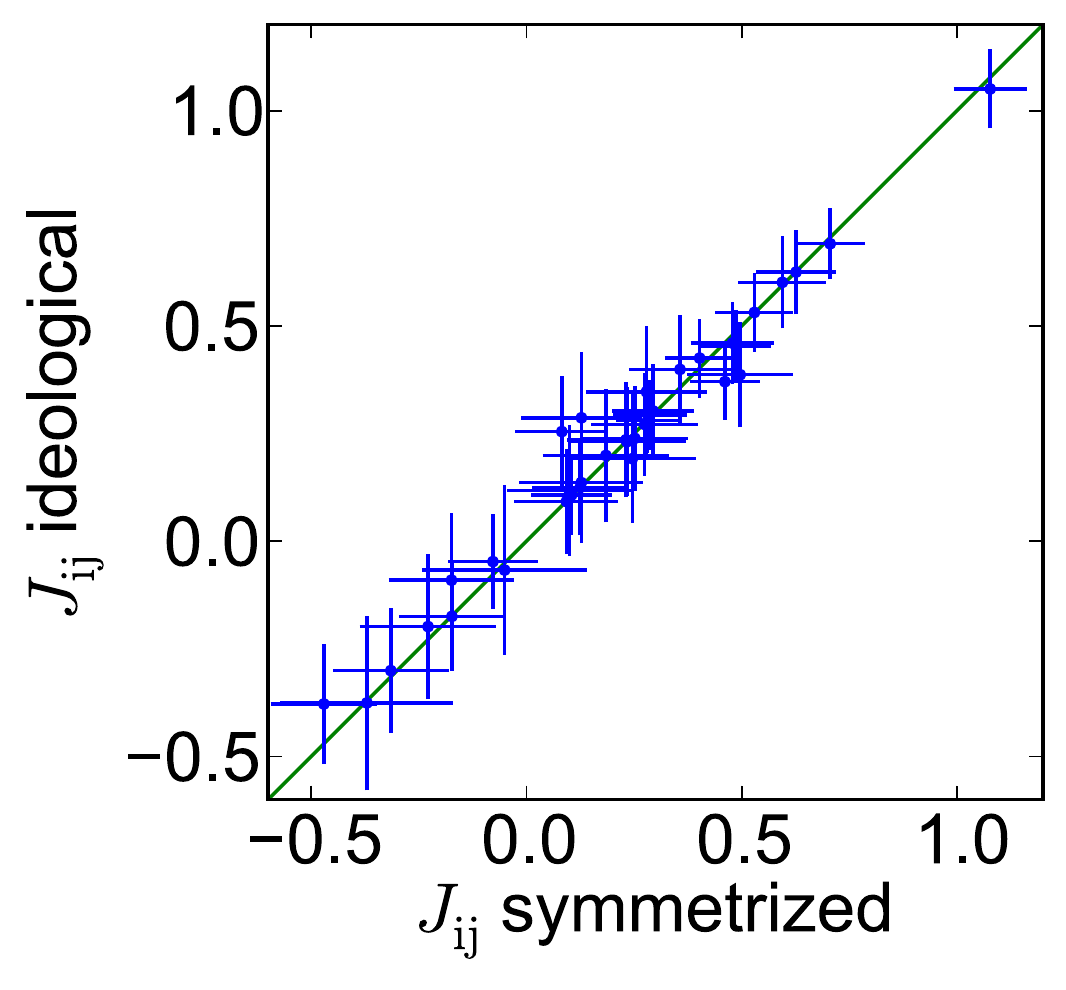}
  \caption{Comparison of couplings from ideological against symmetrized data. The correlation coefficient between average couplings over bootstrap samples is 0.98. Error bars are the standard deviations of our estimates, as in Fig \ref{graph:Jijsigns}A. Green line is 1:1.\label{JvsJ}}
\end{figure}

Returning to the maximum entropy model, we have now the energy function of Eq (\ref{boltzA3}),
\begin{align}
    E(\s) = -\sum_\mr{i} h_\mr{i} \s_\mr{i} - \frac{1}{2}\sum_\mr{ij} J_\mr{ij} \s_\mr{i}\s_\mr{j} .\label{eq: E ising ide}
\end{align}
Solving the model on this data, we find the couplings and mean biases shown in Fig \ref{graph: Cij, Jij ide}B.  If each justice votes with reference to his or her own ideological bias, ignoring the other justices, then we would have $\langle \sigma_{\rm i}\rangle = \tanh(h_{\rm i})$.  In Fig \ref{h_vs_s} we see that this relation does not describe the data well at all.    Thus, even if we account for individual ideological biases, interactions among the justices are still important.  Indeed, if we compare the interactions $J_{\rm ij}$ that emerge from the ideologically labeled votes with those that emerge from the symmetrized data (Fig \ref{Jij}), we see (Fig \ref{JvsJ}) that these are almost the same, across their full dynamic range. Putting the full model together, we see that it again provides a very good account of the data, as shown in Fig \ref{IDE_comp}.  Computing entropies as in Appendix E, we find that this model captures a fraction  $F = 0.92 \pm 0.03$ of the multi--information.

\begin{figure*}[bt]\centering
  \includegraphics[width=\textwidth]{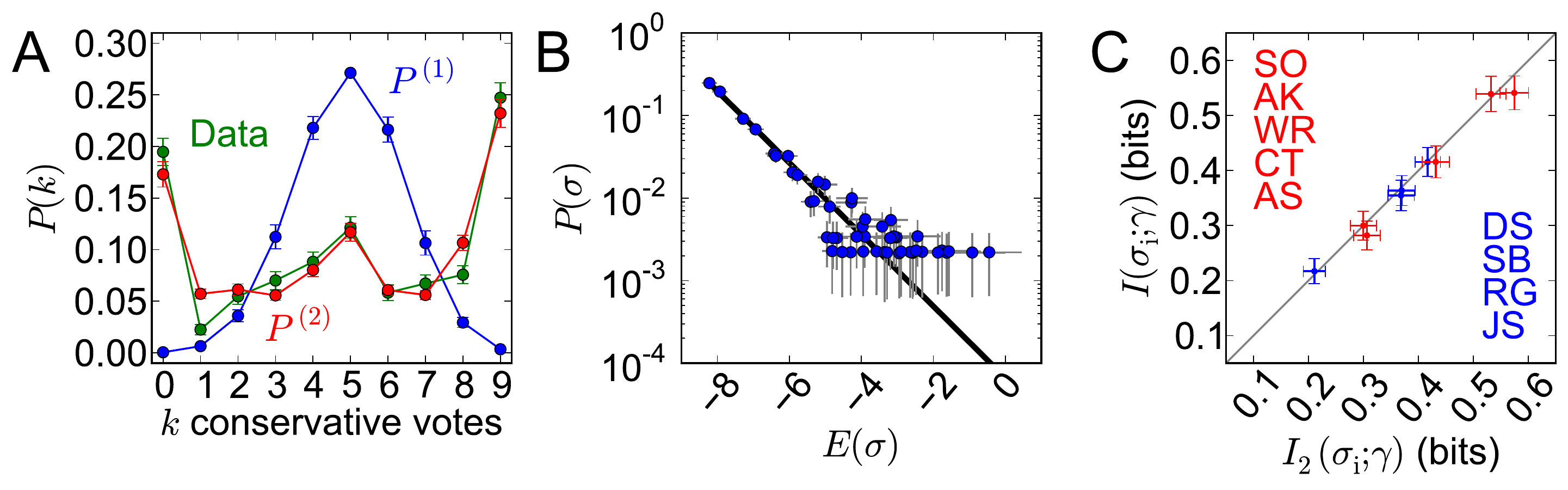}
  \caption{Testing the ideological, maximum entropy model for the Rehnquist court. (A) Probability of $k$ conservative votes. We compare the data (green) with the predictions of the pairwise maximum entropy $P^{(2)}$ (red), and with a model of independent votes $P^{(1)}$ (blue). (B) Probability of each of the 128 observed voting patterns vs the ``energy'' in Eq (\ref{eq: E ising ide}); line is Eq (\ref{ising1}).  Errors as in Fig \ref{fit}. Only states that appear more than once are shown. (C) Mutual information $I(\sigma_{\rm i};\gamma)$ between individual votes $\sigma_{\rm i}$ and the decision $\gamma$ of the majority.  Conservatives are red and liberals blue, from highest $I(\sigma_{\rm i};\gamma)$ to lowest according to data. Error bars represent standard deviations.}\label{IDE_comp}
\end{figure*}

Several points about Fig \ref{IDE_comp} seem worth noting, especially in relation to the corresponding Fig \ref{fit} which shows the quality of model predictions in the symmetrized data.  First, it is clear that the model correctly predicts the emergence of consensus on issues that favor both conservative and liberal positions  (Fig \ref{IDE_comp}A); these consensus votes would be incredibly unlikely if each justice followed his or her biases independently.  The probability of each observed voting pattern is predicted, with essentially the same accuracy as in the symmetrized case (Fig \ref{IDE_comp}B), and the maximum entropy model again captures very precisely the correlation between individual justices and the court majority (Fig \ref{IDE_comp}C).

If we map the energy landscape in the case of ideologically labeled votes, we see slight but significant differences  from the symmetrized case. We still have the largest valleys around the unanimous votes, but the conservative basin has $\sim 25\%$ more weight, as can be seen from Fig \ref{IDE_comp}A.    There is a valley surrounding the 5--4 split, with conservatives in the majority, and a second smaller valley around a 5--4 split with conservatives voting liberally.  We still see the valley around the 7--2 vote against AS and CT, but only one such valley exists since these two justices are so reliably conservative.

While the ideologically labeled data has somewhat more structure than the symmetrized data, it seems fair to summarize our  analysis by saying that keeping  track of the ideological biases of the justices in relation to the content of the question before the court adds relatively little to our predictive power.  How is this possible?

The essential feature of a maximum entropy model is the predicted energy landscape.  For the symmetrized model, this landscape has multiple valleys, corresponding to unanimous votes and 5--4 ideological splits, as well as the smaller valleys in which AS and CT dissent from their seven colleagues (Fig \ref{E_proj}).  This organization emerges collectively from the interactions among the judges, and we have seen that these interactions encode the ideological differences on the court even though we did not introduce these explicitly in constructing the model.  Once the court is ``polarized'' along ideological lines, it takes only very small biases to align the the polarized vote with the right/left content of the question before the court.

One of the main intuitions behind the use of statistical physics ideas in the description of social dynamics is that the emergence of consensus or polarization is analogous to the emergence of order in physical systems at thermal equilibrium: having everyone in a group agree to vote the same way reminds us of all the spins in a magnet ``agreeing'' to point in the same direction.  Importantly, once all the spins in a magnet agree to point in the same direction, even a very small external magnetic field is sufficient to get the entire magnet pointing north.  Concretely, the energy difference between a single electron spin pointing up or down in the earth's magnetic field is much, much smaller than the energy $k_B T$ that sets the scale of random thermal motion:  individual spins {\em do not} point north reliably, although the collective magnetization of a compass magnet certainly does.  Similarly, the biases which couple individual justices' ideological preferences to the merits of individual cases are weak, insufficient to induce unanimity or even to predict correctly the probability of a 5--4 split.  What we see in the patterns of Supreme Court votes is dominated by the emergence of collective states, which then align to the particulars of individual cases.  This is not a metaphor or analogy, but rather the description of a precise, quantitative model that predicts almost all the structure of these votes from the pattern of pairwise correlations.


\begin{thebibliography}{99}
\bibitem{notes}
This paper is written for a physics audience, and thus assumes familiarity with the relevant ideas from statistical mechanics.  In several Appendices, we try not only to expand on technical matters but also to give some background and discussion that we hope will make the work accessible to a broader audience.
\bibitem{ideology}
We use ideological labels from \cite{data} and exclude cases that lack these labels. There is some disagreement---but not much---about the trustworthiness of the labels.
\bibitem{connect}
This can be seen as part of a larger effort to use the maximum entropy principle as a way of building statistical mechanics models of complex biological systems directly from data.  See, as recent examples, Refs \cite{bialek+al_12,tkacik+al_13}, and references therein.
\end{thebibliography}
\end{document}